\documentclass[acmlarge,screen]{acmart}

\usepackage{adjustbox}
\usepackage{multirow}
\usepackage{booktabs,subcaption,dcolumn}
\usepackage{tikz}
\usepackage{pifont}
\usepackage{xcolor}
\usepackage{color}
\usepackage{soul}

\usepackage{wrapfig}
\usepackage{amsmath,amsfonts}
\usepackage{algorithmic}
\usepackage{graphicx}
\usepackage{textcomp}

\usepackage{multirow}
\usepackage{rotating}
\usepackage{makecell}
\usepackage[english]{babel}
\usepackage[font=scriptsize]{caption}
\usepackage{footnote}
\usepackage{listings} 
\usepackage{pifont}

\usepackage{xspace}
\usepackage{colortbl}

\newcolumntype{?}{!{\vrule width 1.5pt}}

\newcommand\revision[1]{%
  \bgroup
  \hskip0pt\color{black}%
  #1%
  \egroup
}

\newcommand{\recommendation}[1]{
    \vspace{1em}
    \noindent\fbox{%
        \parbox{\columnwidth}{%
            \textbf{Recommendation}: {#1}
        }%
    }
    
}

\newcommand{\montimage}{Montimage}
\newcommand{\mmt}{MMT}
\newcommand{\sgrupo}{S2Grupo}
\newcommand{\caiac}{CAIAC}

\newcommand{\takeaway}[1]{
    \vspace{1em}
    \noindent\fbox{%
        \parbox{\columnwidth}{%
            \textbf{Takeaway}. {#1}
        }%
    }
    
}

\setcopyright{acmcopyright}

\AtBeginDocument{%
  \providecommand\BibTeX{{%
    \normalfont B\kern-0.5em{\scshape i\kern-0.25em b}\kern-0.8em\TeX}}}

\setcopyright{acmcopyright}
\copyrightyear{2022}
\acmYear{2022}
\acmDOI{x}

\begin{document}
\title{The Role of Machine Learning in Cybersecurity}

\author{Giovanni Apruzzese}\authornotemark[1]
\orcid{0000-0002-6890-9611}
\email{giovanni.apruzzese@uni.li}
\author{Pavel Laskov}\authornotemark[1]
\email{pavel.laskov@uni.li}
\affiliation{%
  \institution{University of Liechtenstein}
  \city{Vaduz}
  \country{Liechtenstein}
}

\author{Edgardo Montes de Oca}
\email{edgardo.montesdeoca@montimage.com}
\author{Wissam Mallouli}
\email{wissam.mallouli@montimage.com}
\affiliation{%
  \institution{Montimage}
  \country{France}
}

\author{Luis Búrdalo Rapa}
\email{luis.burdalo@s2grupo.es}
\affiliation{%
  \institution{S2 Grupo}
  \country{Spain}
}

\author{Athanasios Vasileios Grammatopoulos}
\email{gramthanos@gmail.com}
\author{Fabio Di Franco}
\email{fabio.difranco@enisa.europa.eu}
\affiliation{%
  \institution{ENISA}
  \country{Greece}
}

\begin{abstract}
Machine Learning (ML) represents a pivotal technology for current and future information systems, and many domains already leverage the capabilities of ML.
However, deployment of ML in cybersecurity is still at an early stage, revealing a significant discrepancy between research and practice. Such discrepancy has its root cause in the current state-of-the-art, which does not allow to identify the role of ML in cybersecurity. The full potential of ML will never be unleashed unless its pros and cons are understood by a broad audience.

This paper is the first attempt to provide a holistic understanding of the role of ML in the entire cybersecurity domain---to any potential reader with an interest in this topic. We highlight the advantages of ML with respect to human-driven detection methods, as well as the additional tasks that can be addressed by ML in cybersecurity. Moreover, we elucidate various intrinsic problems affecting real ML deployments in cybersecurity. Finally, we present how various stakeholders can contribute to future developments of ML in cybersecurity, which is essential for further progress in this field. Our contributions are complemented with two \textit{real} case studies describing industrial applications of ML as defense against cyber-threats.
\end{abstract}

%
%

\ccsdesc[500]{Security and privacy}
\ccsdesc[500]{Computing methodologies~Machine learning}

\keywords{Cybersecurity, Incident Detection, Machine Learning, Artificial Intelligence}

\maketitle

\section{Introduction}
\label{sec:intro}
With the rising complexity of modern information systems and the resulting ever increasing flow of big data, the benefits of Artificial Intelligence (AI) are now widely recognized. Specifically, Machine Learning (ML) methods~\cite{jordan2015machine} are already deployed to solve diverse real world tasks---especially with the advent of deep learning~\cite{lecun2015deep}. Fascinating examples of practical achievements of ML are machine translation \cite{volkart2018statistical}, travel and vacation recommendations \cite{haldar2019applying}, object detection and tracking \cite{ramanathan2021predet} and even various applications in healthcare~\cite{dayan2021federated}.
Furthermore, ML is rightly considered to be a technology enabler, as it has shown great potential in the context of telecommunication systems~\cite{morocho2019machine} or autonomous driving~\cite{al2017deep}. 

Nevertheless, modern society is increasingly relying on information technology (IT) systems---including autonomous ones---which are also actively leveraged by malicious entities. Digital threats are, in fact, continuously evolving~\cite{kettani2019threats}, and according to Gartner attackers will have sufficient capabilities to \textit{harm or kill} humans by 2025~\cite{gartner2021kill}. 
To prevent such incidents and mitigate the plethora of risks that can target current and future IT systems, defensive mechanisms require the capability to quickly adapt to the (i) mutating environments and (ii) dynamic threat landscape.

Coping with such twofold requirement via static and human-defined methods is clearly unfeasible, and deployment of ML in cybersecurity is inescapable. Not surprisingly, abundant work addressed integration of ML in cybersecurity, as evidenced by recent survey papers (e.g.,~\cite{berman2019survey, apruzzese2018deep, gardiner2016security}) and technical reports (e.g.,~\cite{bellasio2020impact, enisa2020threat}). Despite impressive results in research settings, however, the development and integration of ML in production environments is progressing at a slow pace. A recent survey~\cite{kshetri2021economics} shows that although over 90\% of companies already use some AI/ML in their defensive tools, we observe that most of these solutions still leverage `unsupervised' methods (e.g.,~\cite{Darktrace:ML, Lastline:AI}) and mostly for `anomaly detection'. Such observation demonstrates a drastic discrepancy between research and practice, especially in comparison with other domains where ML has already become an indispensable asset. 

The peculiarity of the security domain is that all operational decisions---made by the top management---are about the trade-off between losses and losses~\cite{jalali2019decision}. In simple terms, the rationale is ``paying $x$ to avoid paying $y \gg x$''. Investment in security should be justified by the prevention of substantially higher \textit{but ultimately unpredictable} losses from security incidents. Hence, \textit{decision makers} must have a clear understanding of the (i) benefits, (ii) problems, and (iii) challenges of a cybersecurity solution before endorsing their adoption in practice. However, the current state-of-the-art of ML for cybersecurity fails to deliver such understanding. Taken individually, research papers---commonly claiming to outperform previous work---often lead to contradictory results. For instance,~\cite{vinayakumar2019deep} show that deep learning methods outperform `traditional' ML methods, but the opposite is claimed by~\cite{pontes2021new} in the exact same setting. 
Furthermore, existing literature surveys related to ML in cybersecurity do not provide a holistic coverage suitable for operational decisions. Some of them are too technical and hence tailored for ML experts (e.g.,~\cite{yinka2020review}), others focus only on research efforts neglecting real world implications (e.g.,~\cite{apruzzese2018deep}) or have a limited scope (e.g., only deep learning~\cite{berman2019survey}). 
As a result, the role of ML in cybersecurity is portrayed in a highly fragmented way, thus hindering deployment of ML in practice---despite its great potential for cybersecurity. 

We attempt to rectify this problem. Specifically, this paper is the first effort to provide a comprehensive analysis of the role of ML in cybersecurity. We distill scientific knowledge and industrial experience related to deployment of ML \textit{within the entire domain of cybersecurity}. One of our goals is to make the current state-of-the-art \textit{understandable to any reader}, irrespective of their prior expertise in cybersecurity or ML. We also take this opportunity to clarify many \textit{misconceptions} related to ML in the context of cybersecurity. We highlight the benefits of using ML in cybersecurity by listing all the tasks where it outperforms or provides novel capabilities with respect to traditional security mechanisms. We also elucidate the intrinsic problems of ML in the cybersecurity context. Such analysis reveals the challenges that require the joint contribution of all relevant stakeholders in order to improve the quality of ML-driven security mechanisms. 

Let us explain how we achieve our objective and outline the structure of our paper, which comprises several self-contained sections.
We begin (§\ref{sec:background}) by introducing the key concepts of the ML paradigm in a \textit{notation-free} form. We also define the intended audience of this paper and outline the differences of our work from previous literature surveys and reports. 

Then, in §\ref{sec:detection}, we present the most emblematic application of ML in security: cyberthreat detection. We distinguish between three broad areas: network intrusion detection, malware detection, and phishing detection, which is common in related literature~\cite{apruzzese2018deep,verma2019data}. The goal of this section is to highlight the \emph{added value} of ML with respect to traditional detection mechanisms.

Next, in §\ref{sec:beyond}, we elucidate the cybersecurity tasks \emph{orthogonal to threat detection} that can exploit the capabilities of ML to analyze unstructured data. In contrast to detection problems which require (costly) labels, raw data is abundant in cybersecurity and can also be exploited via ML. For instance, alerts can be filtered to remove annoying false alarms, or compressed into more manageable reports. Furthermore, information from diverse sources can be cross-correlated to anticipate novel attacks, or to identify the weak-spots of a given organization. The goal of this section is to illustrate that there exist many (and vastly unexplored) additional areas in which ML can be deployed to enhance the security of modern systems.

We continue (§\ref{sec:issues}) by emphasizing the intrinsic problems of cybersecurity applications of ML. Some of these problems (e.g., concept drift, adversarial examples, confidentiality) are fundamental and arise from the contrasting assumptions of cybersecurity and ML. Further problems are specific to either in-house development (e.g., hidden maintenance costs) or commercial products (e.g., limited scope and transparency). The goal of this section is highlighting that ML is not perfect and \textit{real} deployments involve many tradeoffs, which must be known (to decision makers), mitigated (by ML engineers), and addressed (in future work).

As our main constructive contribution, we outline the impending challenges of ML in cybersecurity in §\ref{sec:future}. 
Solving these challenges will strongly facilitate the operational deployment of ML in cybersecurity. However, it requires the joint effort of (i) regulatory bodies, (ii) corporate executives, (iii) ML engineers and practitioners, and (iv) the scientific community. Our takeaway is that rectifying the current immaturity of ML in cybersecurity requires \textit{a radical re-thinking} of future technological developments. For instance, research efforts should focus on more pragmatic results instead of merely ``outperforming the state-of-the-art''. However, such efforts necessitate an increased availability of real data whose disclosure requires authorization by senior management, as well as potentially new regulations that enable public release of such data.

To establish a connection between research and practice, we discuss two \textit{real} industrial applications of ML in cybersecurity in §\ref{sec:casestudies}. We note that commercial security products are typically provided as `black boxes' with little technical details about the actual implementation of ML. This section sheds light into the operational tradeoffs and `tricks of the trade' needed to meet the practical needs of the customers. These case studies are provided with the contribution of \montimage{} and \sgrupo{}\footnote{The names of such vendors are anonymized for Double Blind submission policies.}.

This paper is a result of collaboration between researchers, industry practitioners and policy-makers. Our findings reflect the insights from both recent technical reports and scientific literature. 
To the best of our knowledge, no previous work combines such a broad scope with our heterogeneous intended audience.

\textbf{Contribution}
Our main goal is to foster the deployment of ML in cybersecurity by bridging the gap between research and practice. Specifically, our paper make the following contributions:
\begin{itemize}
    \item it provides an overview of the \textit{benefits} and \textit{problems} of ML in the \textit{entire} cybersecurity domain;
    \item it considers the twofold perspective of the \textit{research} and \textit{industrial} community;
    \item it identifies many \textit{misconceptions} that are becoming common in this field;
    \item it highlights how (i) regulatory bodies, (ii) corporate executives, (iii) engineers, and (iv) the research community can contribute to \textit{future developments} of ML in cybersecurity.
    \item it elucidates two \textit{real deployments} of ML products.
\end{itemize}
Furthermore, this paper is meant to be understandable by \textit{any} reader, irrespective of their technical expertise.

\section{Background and Motivation}
\label{sec:background}

To set-up the stage for our paper, we first introduce the main concepts of Machine Learning in a simplified way, accessible to any reader (§\ref{ssec:glossary}). Our goal is to present the established terminology as well as the common classes of existing ML methods. We then define the scope and target audience of this paper (§\ref{ssec:scope}), and highlight the differences of our effort with respect to previous work (§\ref{ssec:related}).

\subsection{Soft Introduction to Machine Learning} 
\label{ssec:glossary}

\begin{figure}[!htbp]
    \centering
    \includegraphics[width=1\columnwidth]{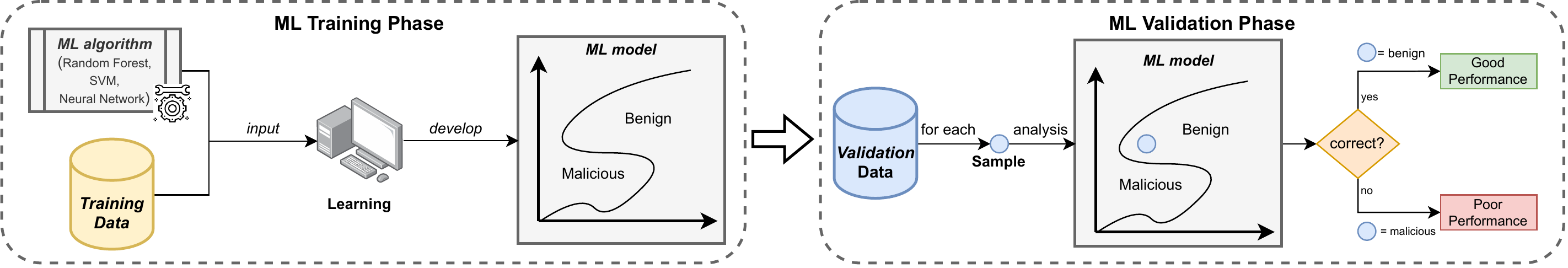}
    \caption{\textit{Machine Learning development}. After collecting some training data and analyzing such data via a ML algorithm, a ML model is obtained. Such ML model must be tested via some validation data. If the performance of such assessment is appreciable, then the ML model can be deployed in production.}
    \label{fig:ml}
\end{figure}

The goal of Machine Learning (ML) is to develop machines\footnote{The notion of a `machine' refers to a software component that can be deployed on any computing device, even in the cloud.} that automatically learn to make decisions. The learning is done through a \textit{training} phase: by instructing a computing device to analyze some `existing' (training) data via a given ML \textit{algorithm}, a ML \textit{model} is developed. Such a model incorporates all the knowledge learned during the training phase, and implements a function to make decisions on `future' data. Before a ML model can be deployed in an operational environment, its performance must be assessed. To this end, some `validation' data is processed by the ML model and its predictions are either analyzed by humans or compared with some known ground truth. We can hence define a ML \textit{method} as ``the process for developing a ML model by using ML algorithms on some training data''. An exemplary workflow of the training and validation phases is schematically depicted in Fig~\ref{fig:ml}.

A crucial factor in the development of ML models is the notion of \textit{labels}, which represent the target value for a prediction function on a given sample (e.g., benign or malicious). Depending on the availability of labels, ML methods can be classified into \textit{supervised} and \textit{unsupervised}. Supervised methods explicitly require labelled training data. In some cases, such labels occur naturally\footnote{In time-series forecasting~\cite{bontempi2012machine}, the learning is done by analyzing the past history of a given phenomenon, which is used to make the future predictions. Such history can be seen as the training data, where each element is associated to its timestamp and its known value (i.e., the label).}, otherwise acquiring labels involves dedicated manual verification. On the other hand, unsupervised methods either do not require labels, or involve limited supervision. For instance, in \textit{reinforcement learning} the ML model is built through a feedback mechanism\footnote{Such mechanism only requires defining the `actions' that can be taken by the ML model, and the `reward' that should be provided to the ML model depending on the effects of its actions on the `environment'.} that is completely automated. 

An orthogonal classification of ML methods is between \emph{shallow} and \emph{deep} learning. Deep learning refers to ML methods based on \textit{neural networks}, which typically require more computational power and larger training datasets compared to shallow ML methods---requirements that could only be met in the recent years~\cite{lecun2015deep}.
Let us point out the first \textbf{misconception}: deep learning is not not necessarily better than shallow ML. Indeed, when the data to analyze has a small number of features, shallow ML can attain similar performance as deep learning~\cite{apruzzese2018deep}, but the latter still requires more resources and the results are more difficult to interpret (e.g.,~\cite{amarasinghe2018toward}).
In contrast, the advantages of deep learning lie in its ability to deal with data with high complexity, such as images, unstructured text, or when temporal dependencies must be taken into account. In all such cases, shallow ML simply cannot be used. Deep learning methods can be supervised or unsupervised, and can also leverage reinforcement learning---e.g., the popular generative adversarial networks (GAN)~\cite{anderson2016deepdga}.

\begin{figure}[!htbp]
    \centering
    \includegraphics[width=0.5\columnwidth]{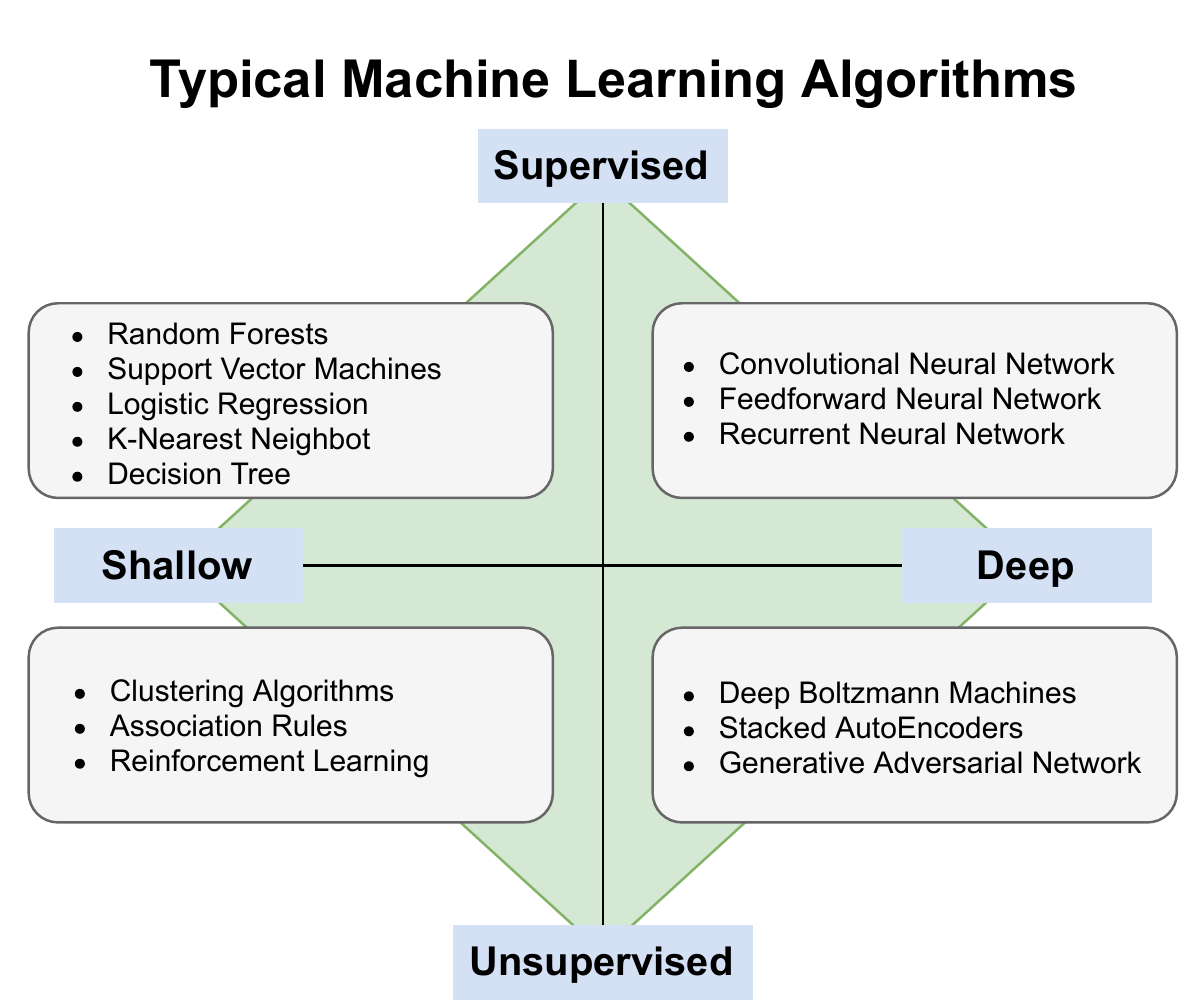}
    \caption{Typical Machine Learning algorithms. An algorithm can be `deep' if it relies on \textit{neural networks}, otherwise it is `shallow'. Algorithms requiring \textit{labelled} data are used for `supervised' tasks, otherwise they can be used also in `unsupervised' tasks.}
    \label{fig:algorithms}
\end{figure}

We provide an overview of some of the most popular ML algorithms in the above mentioned categories in Fig.~\ref{fig:algorithms}. For a more comprehensive description, we refer the reader to~\cite{apruzzese2018deep}.

Finally, let us briefly address the performance assessment of ML models. The most common quality measure is the \textit{accuracy} metric, which represents the percentage of correct predictions made by the ML model. However, \textbf{accuracy can be misleading} in the presence of imbalanced data distributions, which is typical in cyber-threat detection because malicious activities tend to be rare events and are (hopefully) overshadowed by benign samples. In such context, it is common to differentiate between `positives' (i.e., malicious activities) and `negatives' (i.e., benign activities). The performance can then be measured by taking into account the correct (i.e., True Positives and True Negatives) and incorrect (i.e., False Positives and False Negatives) predictions generated by a given ML model. 
A complete list of the most common performance metrics is in Table~\ref{tab:metrics}. Note that performance assessment pertains to ML \emph{models} and not methods. Depending on the specific setting---e.g., the training data, the ML algorithm, its parameters---a ML method may yield many ML models, each having a different performance.

\begin{table*}[htbp]
\centering
\caption{Typical ML performance metrics. Cyber-threat detection represents a binary classification problem: samples are positive (malicious) or negative (benign). Accuracy is useful for cybersecurity tasks where such distinction is not possible. Acronyms: True Positive (TP), True Negative (TN), False Positive (FP), False Negative (FN).} 
    \resizebox{0.8\columnwidth}{!}{
        \begin{tabular}{c?c?c}
            \toprule
             
             \textbf{Metric} & \textbf{Description} & \textbf{Formula} \\
             \toprule

             \begin{tabular}{c}
                  Accuracy \\
                  ($Acc$)
             \end{tabular} & 
             \begin{tabular}{c}
                  It denotes the percentage of correct predictions among all predictions. \\
                  It is misleading in the presence of imbalanced distributions (common in cybersecurity).
             \end{tabular} &    $Acc = \frac{TP + TN}{TP + TN + FP + FN}$ \\ \midrule

             \begin{tabular}{c}
                  Detection Rate \\
                  ($DR$)
             \end{tabular} & 
             \begin{tabular}{c}
                  It measures the capacity of identifying attacks, but it does not consider false alarms. \\
                  It is also known as \textit{Recall}, or \textit{True Positive Rate}.
             \end{tabular}
              
              &    $DR = \frac{TP}{TP + FN}$ \\ \midrule
             
             \begin{tabular}{c}
                  False Positive Rate \\
                  ($FPR$)
             \end{tabular} & 
             \begin{tabular}{c}
                  It represents the percentage of incorrect `positive' predictions. \\
                  Useful for measuring the amount of false alarms.
             \end{tabular}
              &    $FPR = \frac{FP}{FP+TN}$
     \\ \midrule
             
             \begin{tabular}{c}
                  Precision \\
                  ($Prec$)
             \end{tabular} & 
             \begin{tabular}{c}
                  It denotes the percentage of correct predictions among all `positive' samples. \\
                  High values implies low false positives, but nothing can be said about false negatives.
             \end{tabular}
               &    $Prec = \frac{TP}{TP + FP}$ \\ \midrule

             \begin{tabular}{c}
                  F1-score \\
                  ($F1$)
             \end{tabular} & 
             \begin{tabular}{c}
                  It combines \textit{Precision} and \textit{Detection Rate} into a single metric.  \\
                  Useful for `overviews', but it is difficult to interpret.
             \end{tabular}
             
              &    $F1 = \frac{TP}{TP+0.5(FN+FP)}$ \\
             
            \bottomrule
        
        \end{tabular}
    }
\label{tab:metrics}
\end{table*}

\subsection{Scope and Target Audience}
\label{ssec:scope}
The scope of our paper is bridging the gap between scientific research and operational practice of ML in cybersecurity.
We do so by unifying in a single document the benefits, problems, and future challenges of ML in the entire cybersecurity domain. Our paper is meant to be understandable by any reader that is interested in ML and its relationship with cybersecurity.

Specifically, we address the following three classes of target readers:
\begin{itemize}
    \item \textit{Decision makers (e.g., Corporate Executives, Chief Information Security Officers)} who need to understand the state-of-the-art. This paper should allow more sound decisions on the adoption of ML and its integration into existing systems to enhance the productivity of security operation centers.
    
    \item \textit{Security professionals (e.g., security consultants and administrators, digital forensics experts)} who should understand the operational issues affecting ML applications in cybersecurity. Such understanding is crucial for a reliable operation of such instruments in practice, as well as for transparent marketing and assessment of commercial ML solutions.
    
    \item \textit{Researchers and Engineers} who are interested in devising novel ML solutions for cybersecurity, improving existing ML systems, or mitigating some of their limitations. The open issues and challenges presented in this paper should guide future developments of ML for cybersecurity.
\end{itemize}
The takeaways of this paper leverage the contribution\footnote{The identity of the authors are concealed due to Double Blind submission policies.} and take into account the standpoints of \textit{all} the above-mentioned classes of readers. For example, experienced engineers may be aware of the shortcomings of ML, but they may not know how such issues are received by decision makers. At the same time, security professionals may know how ML is used, but they can benefit from understanding the most significant future developments in this field. 
\subsection{Related Work}
\label{ssec:related}
With the advent and increasing popularity of ML, abundant works proposed ML solutions for diverse cybersecurity tasks, resulting in hundreds of research papers. Such abundance inspired many literature surveys that aggregate or summarize the state-of-the-art. However, most of such studies may provide a detailed analysis but on a single application, such as cyber risk assessment~\cite{radanliev2020artificial} or IoT security~\cite{bout2021machine}. Others may focus on a specific cyber detection problem, e.g., malware~\cite{ucci2019survey,gardiner2016security,alzahrani2019review}, spam~\cite{karim2019comprehensive, gangavarapu2020applicability} or intrusion detection~\cite{buczak2015survey,liu2019machine}. Some papers do not explicitly focus on ML (e.g.,~\cite{husak2020sok}), whereas others do not focus on cybersecurity (e.g.,~\cite{hu2021artificial, papernot2018sok}).
Finally, many works only consider specific ML paradigms, such as generative adversarial networks (e.g.,~\cite{yinka2020review}), adversarial ML (e.g.,~\cite{martins2020adversarial, apruzzese2019addressing}), reinforcement learning~\cite{nguyen2021deep}, or deep learning~\cite{berman2019survey}---the latter of which is not necessarily the best `universal' ML solution for cybersecurity. Such finding was shown in the well-known work by Apruzzese et al.~\cite{apruzzese2018deep}, which has a more limited scope than our paper because they (i)~focus on the separation between shallow and deep learning, (ii)~do not delve into cybersecurity tasks beyond threat detection, and (iii)~only consider scientific works. Indeed, ML has undergone significant advances in cybersecurity since the publication of~\cite{apruzzese2018deep}---as we will show in our study.

All these papers, while being useful for interested and experienced researchers, cannot be appreciated simultaneously by security specialists, executives and stakeholders---which are included in our target audience. Indeed, the excessive depth or limited scope of prior work does not allow to grasp the true role (current and future) played by ML in the entire cybersecurity domain. 
At the same time, technical reports (e.g.,~\cite{enisa2020threat, bellasio2020impact}) may be easier to understand by security personnel, but are not useful for researchers due to lack of comprehensive guidelines, and do not provide much insight on \textit{real} ML deployments.

We aim to close the gap between research and practice of ML in cybersecurity with a single document. To this end, we shape this paper so that it is understandable---and usable---by any reader, regardless of their technical competence in ML.
With respect to past works, this paper represents a `meta-review' of the state-of-the-art\footnote{We observe that our paper includes almost 200 referenced works. However, most of such works are cited only once, i.e., in the section devoted to the specific problem addressed by the referenced paper.} which provides a (i) \textit{comprehensive} overview and (ii) \textit{practical} recommendations and research directions (iii)~within the \textit{entire} cybersecurity sphere. Moreover, we (iv) clear many misconceptions that are becoming prevalent in this domain. Finally, we (v) address \textit{all} potential stakeholders---which include but are not limited to researchers.
To the best of our knowledge, no existing paper unifies all of the above in a single contribution.
\section{Machine Learning for Threat Detection}
\label{sec:detection}
The security lifecycle spans over three processes: prevention, detection, reaction~\cite{williams2018engineering}. The complete prevention of any cyber threat is recognized as an impossible task, whereas the reaction phase assumes that the damage has already taken place. Hence, most security mechanisms (including ML-based ones) focus on threat \textit{detection}. For instance, it is not possible to prevent the creation of a phishing webpage; however, such threat can be defused by detecting that a given webpage is compromised, and alerting the users before they fall victim to a phishing `hook'.

The detection of cyber threats can leverage two distinct approaches: \textit{misuse}-based and \textit{anomaly}-based. The former, also referred to as \textit{signature}- or \textit{rule}-based, require defining specific `patterns' that correspond to a given threat---under the assumption that future threats will exhibit the same patterns. The latter require defining a notion of `normality', and aim to detect events deviating from such normality---under the assumption that such deviations correspond to security incidents. These two detection approaches are \textit{complementary}: misuse-based approaches are very precise, but can only detect known threats; anomaly-based approaches tend to generate more false alarms, but have a better chance against novel attacks.

Before the advent of ML, detection mechanisms required \textit{manual} definition of all the necessary elements for a given approach (either misuse- and anomaly-based). Aside from being a time consuming and error prone task, such efforts could not cope with the increasing growth of modern environments. Hence, with the progress of data analytics techniques, detection systems began to leverage \textit{data-driven} solutions, such as ML. These solutions not only required less manual effort but, in some cases, even outperformed traditional hand-written detection schemes~\cite{buczak2015survey}. In the context of ML, such increased performance is due to the intrinsic ability of ML to learn `weak' signals---unnoticed by human operators---in the analyzed data, and use such signals to enhance their detection.

\begin{figure}[!htbp]
    \centering
    \includegraphics[width=0.8\columnwidth]{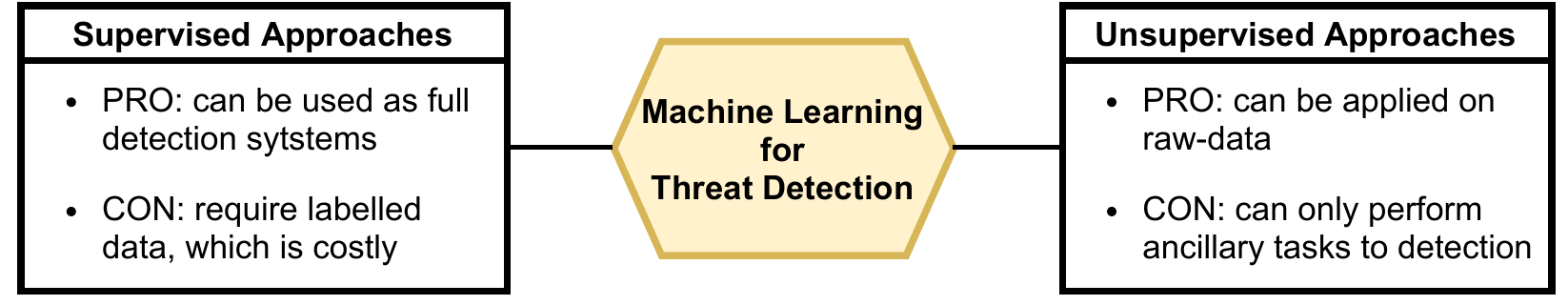}
    \caption{Pros and Cons of Supervised and Unsupervised ML for Cyber Threat Detection.}
    \label{fig:detection}
\end{figure}

The distinguishing characteristic of ML applications for cyber threat detection (schematically depicted in Fig.~\ref{fig:detection}) is whether \textit{supervised} or \textit{unsupervised} ML methods can be deployed. The former can be used as complete detection systems, but require labelled data created via some human supervision. The latter do not have a human in the loop, but can only perform ancillary tasks\footnote{For instance, anomaly detection can be done in an unsupervised fashion, but not all anomalies correspond to security incidents.}. Depending on the data-type to analyze, labels may be easier to acquire: for instance, any layman can distinguish a legitimate webpage from a phishing one~\cite{li2017phishbox}, but distinguishing benign from malicious network traffic is harder~\cite{diaz2020methodology}.

It is common to associate ML methods with anomaly detection (even recent papers suffer from such confusion, e.g.,~\cite{belavagi2016performance}). This is a \textbf{misconception}, because ML can be used also for misuse-based approaches~\cite{khan2019misuse}. Specifically, by analyzing large amounts of data, ML methods can learn the patterns differentiating benign events from malicious ones, so as to automatically define the `signatures' for misuse-based approaches. At the same time, ML can be used for anomaly detection by automatically identifying the `normal' activities that correspond to regular behaviors within a given environment.

Let us elucidate some successful applications of ML aimed at the detection of illicit activities that may occur in a modern enterprise. 
Without loss of generality, we organize this section by distinguishing three broad cyber detection areas: \textit{network intrusion detection} (§\ref{ssec:network}); \textit{malware detection} (§\ref{ssec:malware}); and \textit{phishing detection} (§\ref{ssec:phishing}). 
There are hundreds of works proposing ML for these tasks, and analyzing all such proposals is outside our scope. Hence, we focus on some interesting and recent applications of ML, emphasizing their \textit{practical} results. Our case-studies in §\ref{sec:casestudies} will consider two exemplary applications of ML for cyberthreat detection.

\subsection{Machine Learning in Network Intrusion Detection}
\label{ssec:network}
One of the cybersecurity areas of main interest to modern enterprises is that of Intrusion Detection, which is accomplished by means of Intrusion Detection Systems (IDS). 
An IDS can belong to either of two categories: a \textit{Network} Intrusion Detection System (NIDS) analyzes activities at the network level; whereas a \textit{Host} Intrusion Detection System (HIDS) analyzes activities at the individual host level. In this section we consider NIDS, because HIDS mostly focus on detecting (local) malware which we discuss in §~\ref{ssec:malware}.

\begin{figure}[!htbp]
    \centering
    \includegraphics[width=0.8\columnwidth]{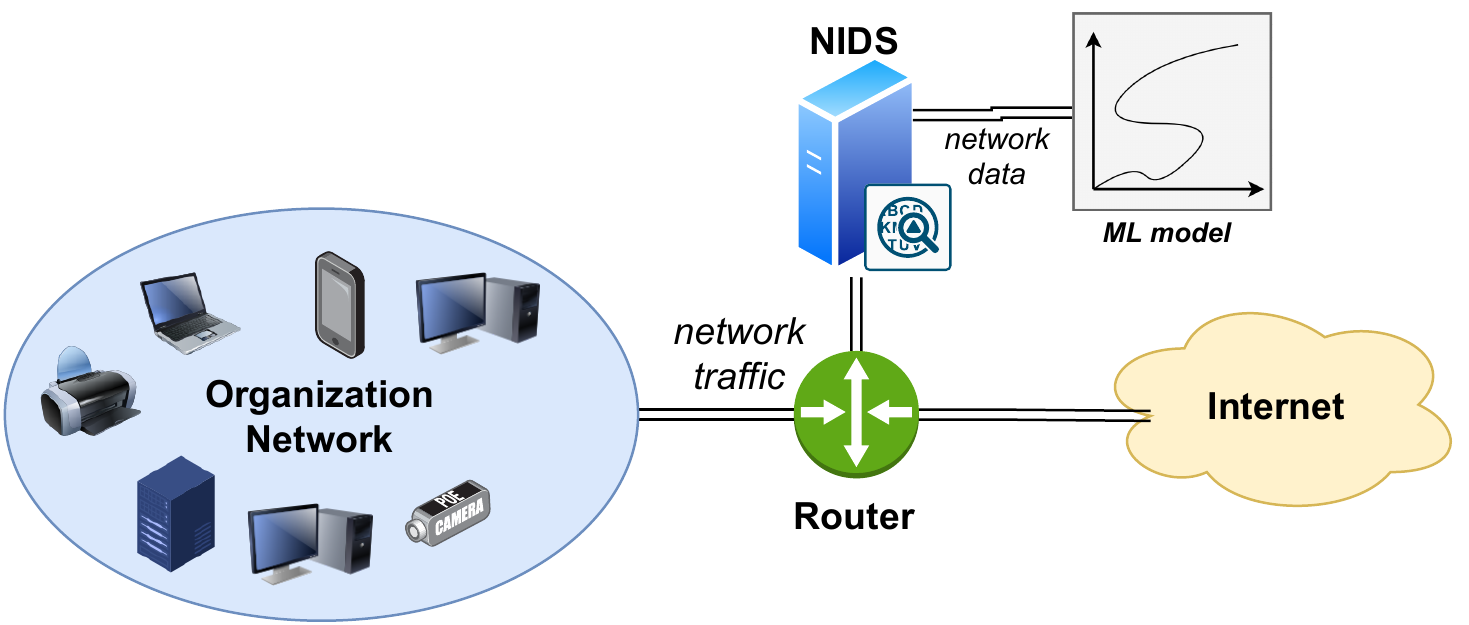}
    \caption{Typical deployment of a ML-NIDS. The border router forwards all the outgoing/incoming network traffic to a NIDS, which further analyzes such data via a ML model.}
    \label{fig:mlnids}
\end{figure}

In the last decade, many ML solutions have been proposed to improve the effectiveness of NIDS, both in scientific literature~\cite{almseidin2017evaluation, buczak2015survey, apruzzese2018deep, berman2019survey}, and in patents (e.g.,~\cite{perdisci2018method, ranjan2014machine}). 
A NIDS can be deployed anywhere in a network environment, and can exploit ML to detect threats against diverse targets, such as cloud, IoT, endpoint devices~\cite{kshetri2021economics}, and even automotive controllers~\cite{lokman2019intrusion}. 
We report in Fig.~\ref{fig:mlnids} the typical deployment of a NIDS that leverages the support of ML, which can analyze data of different types, e.g., full packet-captures (PCAP), network flows\footnote{Netflow: \url{https://www.cisco.com/c/en/us/products/ios-nx-os-software/ios-netflow/}} (NetFlows), SMNP, or even DNS records. Specifically, with the increasing growth of modern networks, NetFlow analyses are preferred due to many advantages over traditional PCAP, such as: reduced privacy concerns, less space required for storage, and faster processing times~\cite{yehezkel2021network}.

ML methods based on \textit{unsupervised learning} are particularly appreciated because acquiring labelled data \textit{for an entire network} is difficult~\cite{diaz2020methodology}. Among these approaches, we highlight the results obtained by \textit{clustering} methods. For example, in~\cite{apruzzese2017periodic} the authors aim to detect attacks by clustering NetFlows with similar temporal behavior and, subsequently, finding the clusters containing hosts that raised alarms from a commercial NIDS based on manual signatures. The results showed a remarkable increase in detection performance\footnote{A similar approach has been successfully integrated even in a commercial product, which we cannot name due to NDA.} with respect to the commercial signature-based NIDS, which only detected 3 malicious hosts, whereas the integration of ML allowed to detect 12. 

Unsupervised methods can also be used to support the (manual) generation of rules for misuse-based NIDS. In CyberProbe~\cite{nappa2014cyberprobe}, the authors cluster honeypot traffic, and create specific rules for each cluster: such rules allowed to detect over $75\%$ attacks which were not included in any security feed. Some papers also exploit unsupervised approaches to counter \textit{lateral movement}\footnote{Lateral Movement: \url{ https://www.lastline.com/blog/lateral-movement-what-it-is-and-how-to-block-it/}}: the approach in~\cite{bohara2017unsupervised} can successfully detect such instances (over $90\%$ recall) with low false positives ($10\%$). Finally, NIDS can also benefit from \textit{deep unsupervised algorithms}. As an example, in Kitsune~\cite{mirsky2018kitsune} the authors use deep learning to analyze PCAP data and improve the detection rate from below 1\% to over 95\% while maintaining a low false positive rate (below $0.1\%$). 
The advantages of unsupervised ML methods make them suitable for commercial products: as an example, the method in~\cite{lombardo2018fast} is used by Aizoon\footnote{\url{https://www.aizoongroup.com/}} to support botnet detection via DNS analyses, achieving less than $0.1\%$ false positive rate. Our detailed case study in §\ref{app:montimage} presents the deployment of unsupervised ML used by \montimage{} to detect anomalous activities in a modern network.

On the other hand, approaches based on \textit{supervised learning}, due to their reliance on good quality labels, are more expensive to deploy but can also provide excellent results. For instance, Exposure~\cite{bilge2011exposure} leverages labelled DNS records to detect domains involved in malicious activities, and achieves less than $10\%$ false alarm rate.
A notable effort against botnets is~\cite{stevanovic2014efficient}, where the authors collect and label some NetFlows, and then use such labelled data to develop a ML botnet detector achieving over $95\%$ precision. Moreover, the work in~\cite{apruzzese2020hardening} proposes the usage of probability labels (instead of binary labels) to detect botnet NetFlows that may evade traditional ML-NIDS, and reach over $97\%$ precision. Remarkable successes also include deep learning methods, such as the approach in~\cite{javaid2016deep} which achieves almost $95\%$ detection rate. 
In particular, we highlight those solutions that combine deep learning with temporal analyses: such twofold perspective allows to detect additional malicious patterns that can improve detection performance. For instance, in~\cite{corsini2021temporal} the F1-score improves from $0.90$ to $0.95$ when also temporal dependencies are considered. We will present a real deployment of a similar solution in §\ref{app:s2grupo}, describing how \sgrupo{} protects Industrial Control Systems (ICS), showcasing the pros (and cons) of ML with respect to older techniques based on heuristics.

Let us conclude with a remark: \textbf{the superiority of deep learning for NIDS is not yet proven}. For instance, the authors of~\cite{pontes2021new} and~\cite{vinayakumar2019deep} both evaluate shallow and deep ML methods on the same dataset (the CICIDS17~\cite{sharafaldin2018toward}): while~\cite{vinayakumar2019deep} claims that deep learning outperforms traditional ML, the authors of~\cite{pontes2021new} achieve the opposite result. Specifically,~\cite{vinayakumar2019deep} shows a `deep' neural network achieving $0.96$ F1-score and a `shallow' decision tree achieving $0.95$ F1-score; whereas~\cite{pontes2021new} show a `deep' neural network also achieving $0.96$ F1-score, but their `shallow' decision tree reaches $0.99$ F1-score. 
Our stance on this subject is that, under the assumption that deep learning is superior, the marginal improvement does not justify its adoption due to its additional complexity and computational requirements.
\subsection{Machine Learning in Malware Detection}
\label{ssec:malware}
The fight against malware is one of the most emblematic challenges of cybersecurity. Because malware affects a specific device, its detection is performed by analyzing data at the host-level, i.e., through HIDS. Indeed, antiviruses can be considered as a subset of HIDS~\cite{kumar2016approach}. 
A given malware variant is tailored for a given operating system (OS). The popularity of Windows OS made it the most common malware target for more than two decades. However, attackers are now turning their attention to mobile devices running, e.g., Android OS\footnote{\url{https://www.gdatasoftware.com/news/2019/07/35228-mobile-malware-report-no-let-up-with-android-malware}}. 

\begin{figure}[!htbp]
    \centering
    \includegraphics[width=1\columnwidth]{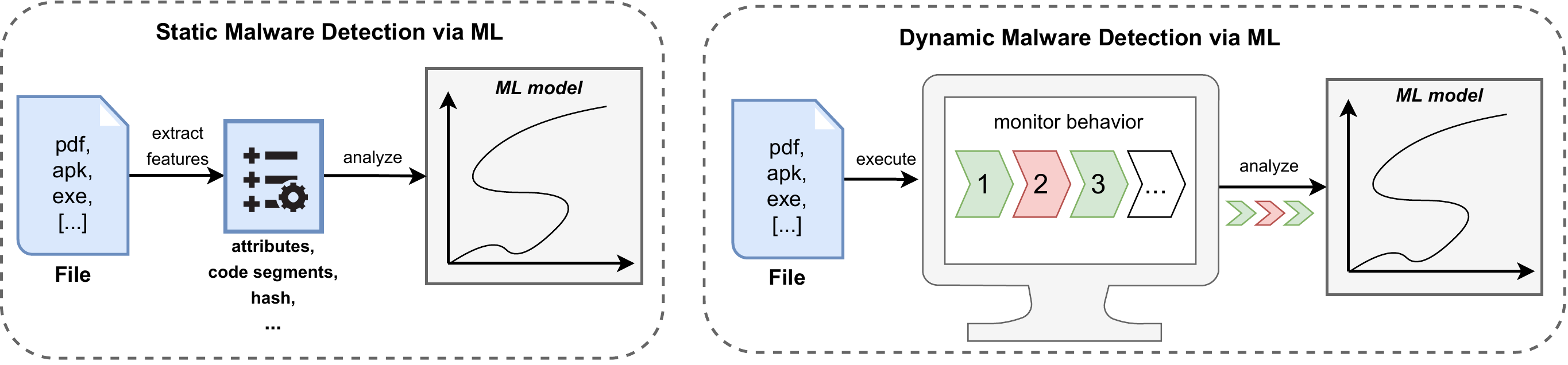}
    \caption{Malware Detection via ML. In static analyses, the properties of a given file are extracted and analyzed by a ML model. In dynamic analyses, the file is executed and the entire behavior is monitored, and then analyzed by a ML model.}
    \label{fig:malware}
\end{figure}

Malware detection can use two types of analyses: \textit{static} or \textit{dynamic}. The former aim to detect malware without running any code, by simply analyzing a given file. The latter focus on analyzing the behavior of a piece of software during its execution, usually by deploying it in a controlled environment and monitoring its activities. Both static and dynamic analyses, schematically depicted in Fig.~\ref{fig:malware}, can benefit from ML.

\paragraph{Static Analysis}
These analyses are simple, particularly effective against known pieces of malware, and can be enhanced via ML in many ways.
For instance, clustering is useful to identify properties of similar pieces of malware. A similar method is proposed in~\cite{hu2013mutantx}, with the goal of finding a common treatment against all elements in each cluster, and reaches up to $90\%$ precision. In contrast, the authors of~\cite{li2017android} leverage clustering to improve the detection of Android malware, and exceed $95\%$ detection rate.
Static analyses can be further improved When labelled data is available. An early example is the detection of malicious PDF files in~\cite{vsrndic2013detection}: here, the authors use ML to analyze the structural properties of PDF files, extracting features that yield proficient detection results (over $99\%$ detection rate with less than $0.001\%$ false positive rate). Recently, a different approach leverages deep learning to transform executables into images, which are then used to perform the detection: the authors of~\cite{kalash2018malware} achieve over $99\%$ accuracy in identifying Windows malware. 

Despite these successes, all static malware detection approaches are prone to evasion. This can be easily achieved by modifying the malware executable, which can be implemented without changing its underlying malicious logic. To aggravate the problem, advanced malware variants (e.g., \textit{polymorphic} or \textit{metamorphic}) automatically modify their executables, defeating any static detection approach. 

\paragraph{Dynamic Analysis}
The combination of dynamic approaches with ML techniques yields effective countermeasures against polymorphic malware. 
Multiple ML solutions exploit clustering: grouping malware with similar behavior allows to focus only on those clusters that have not been seen before. For example,~\cite{rieck2008learning} proposes a dynamic approach combining clustering and anti-virus scanners to detect and sanitize entire groups of malware variants, achieving almost perfect accuracy against Windows malware. More recently, the work in~\cite{amer2020dynamic} focuses on Windows malware by leveraging a combination of graph- and NLP-techniques applied to dynamic API calls, and achieves $99.99\%$ accuracy.
Some papers even propose deep learning, such as~\cite{liu2021research} which uses deep neural networks to extract the most relevant dynamic features to classify Android malware, achieving nearly $80\% accuracy$. Moreover, the authors of~\cite{al2019leveraging} apply deep learning to detect Windows \textit{ransomware}, and achieve $93\%$ detection rate and $97\%$ precision.
An interesting work is HeNet~\cite{chen2018henet}, which leverages ML for dynamic malware detection by analyzing \textit{hardware}-specific (i.e., Intel CPU) data streams, achieving perfect accuracy on real benchmarks.
Finally, it is possible to \textit{combine} static with dynamic analyses via ML: this is done in EC2~\cite{chakraborty2017ec2} which combines unsupervised with supervised ML to detect novel android malware, achieving over $90\%$ detection rate. 

\subsection{Machine Learning in Phishing Detection}
\label{ssec:phishing}

\begin{figure}[!htbp]
    \centering
    \includegraphics[width=1\columnwidth]{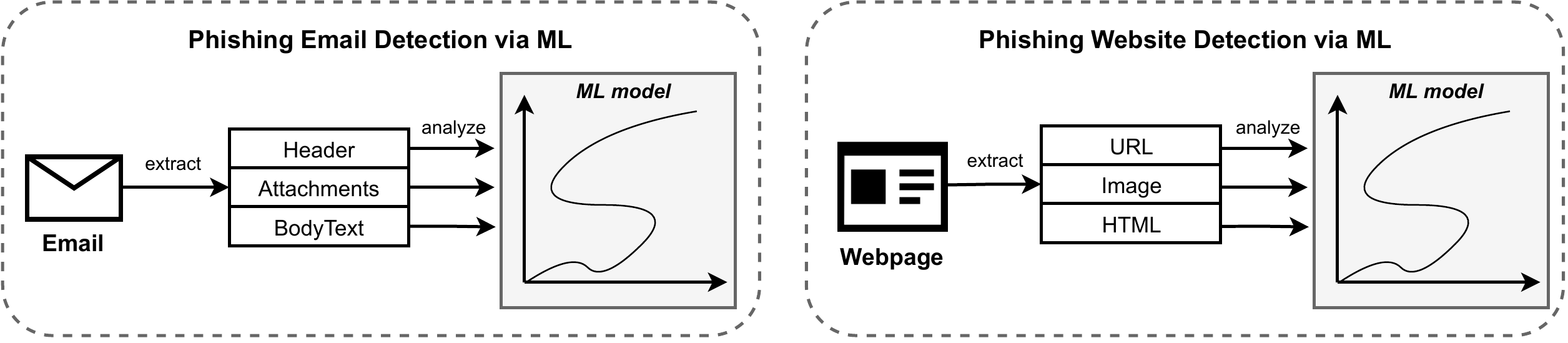}
    \caption{Phishing Detection via ML. For websites, the ML model can analyze the URL, the HTML, or the visual representation of a webpage. For emails, the ML model can analyze the body text, the headers, or the attachment of the email. }
    \label{fig:phishing}
\end{figure}

Phishing represents one of the most common vectors for penetrating a target network and is still a rampant threat in the cybersecurity landscape~\cite{kettani2019threats}. 
Early detection of phishing attempts is of paramount importance to modern organizations, and can greatly benefit from ML. Specifically, we distinguish two different applications of ML to counter phishing attempts: detection of phishing \textit{websites}, where the goal is identifying webpages that are camouflaged to resemble a legitimate website; and detection of phishing \textit{emails}, which either point to a compromised website or induce a response that includes sensitive information. The main difference between these two approaches is the type of analyzed data: for websites, it is common to use the URL of the webpage, its HTML code, or even its visual representation~\cite{tian2018needle}; for emails, it is typical to analyze the text, the header or the attachments of an email~\cite{alhogail2021applying}. A schematic representation of such applications is shown in Fig.~\ref{fig:phishing}, which we now describe in more detail.

\paragraph{Phishing Webpage Detection}
Phishing websites are mostly dealt with via \textit{blacklists}. However, such lists quickly become unreliable because expert adversaries frequently move their phishing hooks from site to site: as shown in~\cite{tian2018needle}, over $90\%$ of `squatting' phishing websites are not detected by popular blacklists. ML represents a viable alternative to manual and static blacklisting, and modern web-browsers already leverage its potential~\cite{liang2016cracking}.

Compared to malware or network intrusion detection, works proposing \textit{unsupervised} ML against phishing websites are less prevalent. An example is~\cite{zhuang2012intelligent}, exploiting clustering to support the detection of phishing websites, and achieving over $95\%$ accuracy. 
In contrast, \textit{supervised} ML is abundant because verifying the legitimacy of a webpage is relatively simple, which facilitates labelling procedures and allows to develop complete ML detectors~\cite{li2017phishbox, tian2018needle, corona2017deltaphish}. 
Some works use ML to analyse features extracted from a given URL. It is interesting to note that while the authors of~\cite{basnet2014learning} use up to 130 features to achieve $99\%$ detection rate, other works (e.g.,~\cite{sahingoz2019machine}) use less than 30 features and achieve similar results.
Other proposals leverage third-party information provided by reputable sources (e.g., DNS records) which can be derived from the URL: an example is PhishMon~\cite{niakanlahiji2018phishmon} which achieves nearly $96\%$ accuracy while maintaining a low $1\%$ rate of false positives.
Some papers consider the twofold perspective provided by the analysis of both URL- and HTML-based features, which is advantageous when the single URL is not enough to identify a webpage as phishing or not. For example, the work in~\cite{babagoli2019heuristic} achieves $95\%$ detection rate by combining these two data types.
A significant work is~\cite{corona2017deltaphish} which combines the inspection of the underlying HTML code of the webpage with image processing techniques (based on deep learning) to identify compromised websites: the results show over $95\%$ detection rate at the cost of $1\%$ false positive rate. Finally,~\cite{tian2018needle} uses all of the above (images, HTML and URL): for nearly 1000 squatting phishing websites, manual blacklisting only detected $9\%$, whereas ML detected $70\%$ of such phishing attempts.

\paragraph{Phishing Email Detection}
One of the earliest applications of ML for cybersecurity involves the detection of unsolicited emails (also often referred to as `spam'). Recent advances in Natural Language Processing (NLP) can be leveraged by ML to analyze the body of an email and identify malicious intent~\cite{biggio2018wild}. 

Only few proposals leverage \textit{unsupervised} ML, such as~\cite{diale2019unsupervised} which achieves over $95\%$ detection rate. However, as it is the case for phishing website detection, acquiring ground truth labels for emails is a trivial task, which facilitates the deployment of \textit{supervised} ML used by email providers to enhance their automatic filters~\cite{karim2019comprehensive}. For instance,~\cite{alhogail2021applying} analyses the text of an email, and reaches almost $99\%$ accuracy with less than $0.01\%$ false positive rate. The authors of Themis~\cite{fang2019phishing} exploit deep learning to analyze both the text and the header of an email and exceed $99\%$ accuracy.
Finally, we mention the work in~\cite{gascon2018reading}, where the authors leverage supervised ML to detect \textit{spear-phishing} attacks by analyzing an email from different perspectives, and achieve over $90\%$ detection rate at the cost of $1\%$ false positives.
Attachments can also be analyzed by any malware detection technique (§\ref{ssec:malware}).

As a small digression, we mention that the fight against phishing (and spam) has recently moved to Online Social Networks. This setting exhibits many similarities with the detection of phishing in emails, as it also involves NLP techniques. As an example, the authors of~\cite{wu2017twitter} use deep learning to detect malicious tweets, and obtain promising results with almost $95\%$ detection rate but with $5\%$ false positive rate. Similarly, MalT$^P$~\cite{lancaster2018maltp} specifically focuses on tweets luring victims to phishing websites, achieving over $95\%$ detection rate and nearly $90\%$ precision.

\takeaway{Using ML for cyberthreat detection has proven to be greatly successful (e.g.,~\cite{mirsky2018kitsune, chen2018henet, tian2018needle}).}

\section{Beyond Detection: additional roles of Machine Learning in Cybersecurity}
\label{sec:beyond}

Besides threat detection, there are many additional roles that ML can cover in cybersecurity.
Indeed, modern environments constantly generate massive amounts of data, which may come from heterogeneous sources---including the very same ML models described in §\ref{sec:detection}. Analyzing such data via (additional) ML can provide insights that further improve the security of digital systems.

\begin{figure}[!htbp]
    \centering
    \includegraphics[width=0.8\columnwidth]{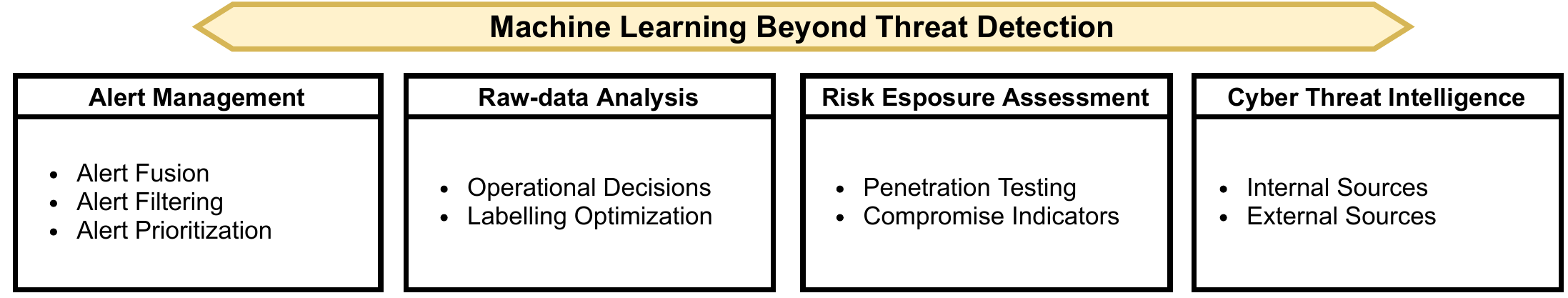}
    \caption{Additional tasks that can be addressed via ML in cybersecurity. All such tasks mostly involve dealing with raw and unstructured data from heterogeneous sources, and provide fertile ground for ML.}
    \label{fig:beyond}
\end{figure}

Without loss of generality, we classify all these complementary roles of ML in four tasks: alert management~(§\ref{ssec:alert}), raw-data analysis (§\ref{ssec:raw}), risk exposure assessment (§\ref{ssec:risk}), and cyber threat intelligence (§\ref{ssec:intelligence}). We now describe each of these tasks, schematically summarized in Fig.~\ref{fig:beyond}.

We highlight an enticing characteristic shared by most ML applications described in this section: they do not require extensive and human-guided labelling procedures, and hence belong to the \textit{unsupervised} ML category. The potential of using raw-data almost `as-is' makes all the ML methods discussed in this section readily applicable in many real scenarios.

\subsection{Alert Management}
\label{ssec:alert}
It is well-known that developing the `perfect' detection system is not possible (with or without ML). Hence, to prevent the automated execution of actions based on wrong predictions, the output of detection systems usually comes in the form of \textit{alerts}. Depending on such alerts (e.g., their relevance, the involved hosts, or their number) a more appropriate response can be taken.
However, modern environments generate thousands of alerts every hour (as shown in, e.g.,~\cite{yagemann2021arcus, apruzzese2017periodic}), making manual triaging an impossible task. To address this problem, ML can be used for \textit{filtering}, \textit{prioritization}, a or even \textit{fusion} of alerts into more general events. 
\begin{itemize}
    \item \textit{Alert Filtering.}
    By definition, an alert is not necessarily malicious, and a significant percentage of alerts correspond to false alarms. Because being notified by many irrelevant alerts is impractical and annoying, ML can help in filtering redundant alerts, e.g., because they are related to the same underlying problem. An example is~\cite{su2019false}, which is specifically tailored for false alarms generated by ML-NIDS: its effectiveness on real botnet traffic is a remarkable reduction of $75\%$ of the time spent on triaging of false alerts, outperforming non-ML mechanisms by $45\%$.
    
    \item \textit{Alert Prioritization.}
    If security administrators face too many alerts, prioritization techniques can be applied to identify the most critical alarms. ML is beneficial as it can automatically `learn' the most relevant ranking criteria with limited supervision. For instance, the very recent work in~\cite{vidovic2021ranking} shows that ML correctly ranks the most sensitive alerts at the top position in $95\%$ of the cases.
    
    \item \textit{Alert Fusion.}
    The most intuitive way to manage large amounts of alerts is to \textit{aggregate} similar alerts, and then to find \textit{correlations} between these groups in order to identify causal relationships relevant for security tasks. For instance, ASSERT~\cite{okutan2019assert} leverages clustering to identify which are the preferred protocols and network ports targeted by malicious activities. Their results highlight that modern attacks are increasingly relying on the Remote Desktop Protocol (RDP), as it enables lateral movement activities through pivoting~\cite{apruzzese2017detection}.
\end{itemize}
All of the techniques above can be \textit{combined} together. In this context, we mention the alert management solution in~\cite{mcelwee2017deep} exploiting deep learning to condense and prioritize alerts: the resulting platform was tested and found usable by real security analysts.

\subsection{Raw-data Analysis}
\label{ssec:raw}
The cybersecurity domain must deal with heterogeneous systems, each generating raw-data of different nature (e.g., logs, reports, alerts). Such setting represents a fertile ground for ML, whose capabilities could be leveraged to maximize the opportunities provided by such raw-data. We can differentiate two areas of application of ML in this context: the support of \textit{operational decisions} via log data analyses; and the use of (unlabelled) data to \textit{optimize labelling} efforts and foster deployment of supervised ML.

\paragraph{Operational Decisions.}
The abundance of \textit{log} data in modern information systems makes ML promising in the context of operational security. The importance of log data analysis became evident after several high-profile security incidents that involved stealthy exfiltration of confidential data\footnote{An example is the well-known RSA incident: \url{https://www.theregister.co.uk/2011/04/04/rsa\_hack\_howdunnit/}.}.
Beehive~\cite{yen2013beehive} was one of the first (unsupervised) ML systems focused on knowledge extraction from heterogeneous log data (generated by proxy, DHCP or VPN servers). The goal was combining all these logs in an anomaly detection fashion: data points not associated with `typical' log patterns represented `incidents' that required manual intervention. Beehive was evaluated on two weeks of log data at the EMC Corporation and detected almost 800 incidents, $65\%$ of which related to true security incidents (malicious activities or policy violations). In comparison, non-ML methods performed much worse, as they were only capable of detecting 8 correct incidents (with a recall of just $1\%$). 
Despite being unsupervised ML, Beehive still required manual feature engineering: the most relevant pieces of information from every log source had to be determined via expert knowledge. Such problem was overcome with the advent of deep learning. A prominent example is DeepLog \cite{du2017deeplog}, which analyzes heterogeneous log data (e.g., Hadoop, or OpenStack logs) with a similar objective as Beehive. DeepLog achieves impressive results in a lab environment, with close to $100\%$ detection rate after training on only $1\%$ of the available data. 

\paragraph{Labelling Optimization}
Many threat detection techniques (§\ref{sec:detection}) rely on supervised ML, which may require huge amounts of labelled data. Such requirement prevents their applicability in real scenarios, because manual labelling can be prohibitive---especially in Network Intrusion Detection. In contrast, unlabelled data is common in cybersecurity, and many efforts proposed \textit{semi-supervised learning} methods to increase the `return' of small sets of labelled data, and hence enable deployment of fully-supervised ML methods~\cite{apruzzese2022sok}.
For instance, the botnet detector in~\cite{zhang2021network} reaches 0.83 F1-score with only 2400 labels; in contrast, the detector in~\cite{apruzzese2020deep} reaches 0.95 F1-score on the same network scenario, but requires millions of labelled samples.
A parallel line of research leverages the so-called \textit{active learning} paradigm. The idea is to use a ML model (trained on a small labelled dataset) to `suggest' which samples should be labelled in a (large) unlabelled dataset, to maximize its `learning rate'. As an example,~\cite{zhang2020enhancing} shows that it is possible to save significant labelling effort (from $30\%$ up to $90\%$) by providing the ground truth of only a restricted amount of samples. An intriguing property of active learning is that it can be used even for already deployed ML models, by following the so-called \textit{lifelong learning} principle: for instance, Tesseract~\cite{pendlebury2019tesseract} can boost its performance from $57\%$ to $70\%$ after being retrained on 700 samples `actively labelled' by a human expert.

\subsection{Risk Exposure Assessment} 
\label{ssec:risk}
Although the complete prevention of any cyber attack is an unreachable objective, a system can be significantly strengthened by focusing on its weak spots and anticipating the most likely threats.
In this context, ML can help for several tasks, such as penetration testing, or estimation of compromise indicators.

\paragraph{Penetration Testing}
By automatically `attacking' existing security systems, ML can be a great asset for vulnerability assessment. For instance, ~\cite{ghanem2018reinforcement} apply \textit{reinforcement learning} to synthetically craft attacks against traditional NIDS: the ML approach found the same amount of vulnerabilities in \textit{half} the time of manual inspection, and achieved a speedup of $90\%$ with respect to a random attack procedures. More recently,~\cite{apruzzese2020deep} adopted a deep reinforcement learning approach to automatically evade, and then harden, a ML-based botnet detector.
Similarly,~\cite{uwagbole2017applied} assessed the vulnerabilities of databases to SQL-injection attacks crafted via ML. There are even proposals of dedicated ML-assisted \textit{platforms} for performing all such assessments~\cite{chaudhary2020automated}. According to a recent survey~\cite{mckinnel2019systematic}, the potential of ML for penetration testing is still vastly unexplored.

\paragraph{Estimation of Compromise Indicators}
It is possible to use ML to estimate the most likely compromised hosts in a given system. The authors of~\cite{yen2014epidemiological} study a corporate environment, using ML to analyze information from heterogeneous sources, such as the behavior of each individual host and of the entire network---as reported by end-point protection devices (McAfee); or even personal information on the specific user of each host. The findings revealed that visits to `business' websites represented the most common indicator of a compromised host (almost $30\%$), with second place for `travel' websites (nearly $15\%$)---this is intriguing, considering that such activities were performed during working hours. A potential opportunity is combining ML with honeypots (with a different scope than~\cite{nappa2014cyberprobe}): such strategy is exploited in~\cite{garre2021novel} to identify which hosts are more likely to be infected by botnet malware.
Finally, Facebook exploits ML to identify fake accounts by correlating different sources~\cite{xu2021deep}, allowing to reduce such annoyance by nearly $30\%$.

\subsection{Threat Intelligence}
\label{ssec:intelligence}
The main task of threat intelligence is to collect and analyze information for \textit{anticipating} novel attacks. This is clearly a powerful instrument for keeping defenses up-to-date in a proactive approach~\cite{biggio2018wild}. However, we observe that a crucial aspect in the protection of enterprises revolves around the value of the items being considered: hence, ML methods for cyber threat intelligence should be configured so as to prioritize the protection of the most business-critical infrastructures. Failure to take this into account may limit the usefulness of ML.

Nevertheless, applications of ML for threat intelligence can leverage either \textit{internal} or \textit{external} data sources (or both).

\paragraph{Internal Sources}
Foreseeing future attack strategies via ML can be done with exclusive reliance on internal corporate data. For instance,~\cite{sweet2020variety} leverages ML to artificially create alerts corresponding to past cyberattacks, and then use such alerts to study an attacker's behaviour---potentially by using additional ML solutions. As an example, SAGE~\cite{nadeem2021alert} exploits ML to compress over 300k individual alerts in less than 100 `attack graphs' representing the specific steps of an entire offensive strategy. 
Another possibility is to use deep learning to `disassemble' some code executables, allowing to identify some potentially malicious patterns that can reappear in future malware: for instance, EKLAVIA~\cite{chua2017neural} achieves a remarkable 80\% accuracy in such task. Finally, internal and external data sources can be mixed: the authors of~\cite{kang2016ensemble} exploit historical malware information (provided by Symantec) to foresee how future malware could affect a corporation, and their ML solution provided up to 4 times as many correct predictions as non-ML baselines.

\paragraph{External Sources}
It is possible to use ML for the so-called open source intelligence (OSINT).
For example, the authors of~\cite{sapienza2017early} focus on security incidents mentioned on Twitter. Their ML approach identified many malicious activities occurring in 2016, such as the Mirai botnet (October 2016) or the data breach at AdultFriendFinder (November 2016), where over 400 million accounts were exposed. Similarly, the deep learning method in~\cite{vinayakumar2019ransomware} analyzed tweets to study the development of ransomware attacks.
It is also possible to use information from security feeds, such as the Common Vulnerability Score (CVS) stored on well-known databases\footnote{An example is the CVE database, storing vulnerabilities as well as their exploitance likelihood: \url{https://cve.mitre.org/}.}. For instance, in~\cite{chen2019vase} the authors use ML to predict the CVS with almost 1 week earlier than traditional cybersecurity feeds. Prediction of the CVS with ML can also be done via darkweb data as shown in~\cite{almukaynizi2017proactive}. The authors use ML to crawl underground forums and correlate meaningful information with vulnerability descriptions. By validating the results via third-party signatures (e.g., Symantec), the proposed ML method successfully predicted the exploitability for about $40\%$ of recorded vulnerabilities compared to about $10\%$ of common feeds. Automated analyses via ML of underground forums (in different languages) aimed at uncovering `cyber-criminal markets' are also performed in~\cite{portnoff2017tools}, allowing to infer the prices of malicious exploits. Finally, we even mention the existence of \textit{patents} that leverage ML to predict cyberattacks in modern environments~\cite{okutan2021cyberattack}.

\takeaway{There are many tasks complementary to threat detection that can be covered by ML. The main challenge lies in obtaining relevant information from unlabelled (e.g.,\cite{apruzzese2022sok, okutan2019assert}) or unstructured data coming from heterogeneous sources (e.g.,~\cite{du2017deeplog, xu2021deep, almukaynizi2017proactive}). Such challenge, however, also represents an intriguing opportunity.}
\section{Intrinsic Problems of Machine Learning in Cybersecurity}
\label{sec:issues}

As shown in §\ref{sec:detection} and §\ref{sec:beyond}, ML can cover a plethora of roles in cybersecurity. 
Yet, in this specific domain, unleashing the full benefits of ML \textit{in practice} is difficult. Such difficulty stems from the underlying conflict between (a) the intrinsic characteristics of the cybersecurity domain, and (b) the fundamental assumptions of ML.

\begin{figure}[!htbp]
    \centering
    \includegraphics[width=0.6\columnwidth]{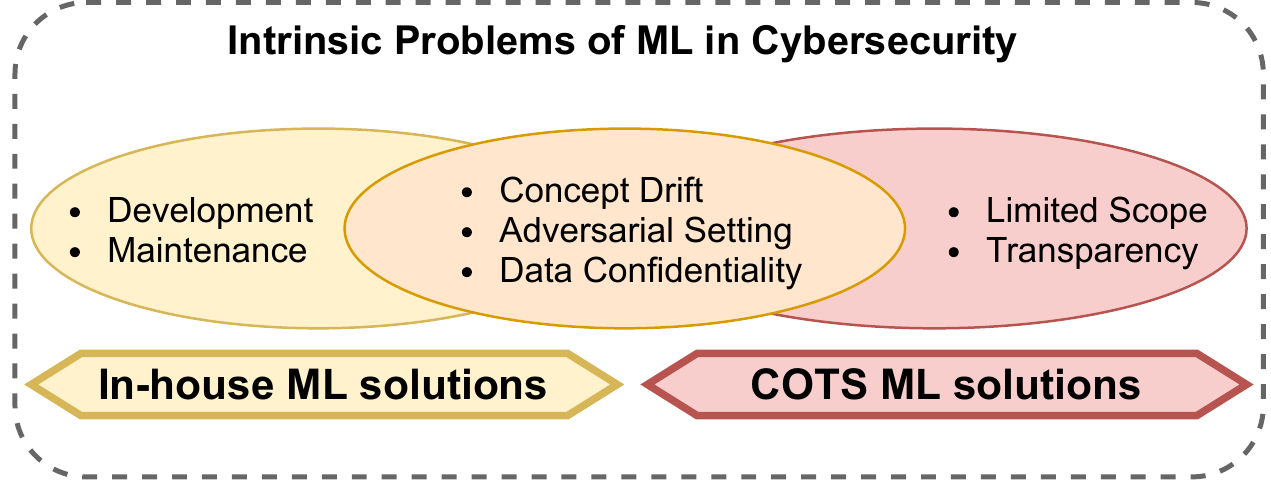}
    \caption{Problems of ML in Cybersecurity. Some are specific to either in-house solutions, or to commercial-off-the-shelf (COTS) ML products. Others are shared by both of these categories.}
    \label{fig:issues}
\end{figure}

Understanding such conflict is crucial for a comprehensive assessment of all the tradeoffs pertaining to ML-based cybersecurity solutions. 
Therefore, we now discuss the \textit{intrinsic problems of ML in cybersecurity}, for which we provide an overview in Fig.~\ref{fig:issues}. Specifically, we begin by presenting the problems affecting \textit{any} ML solution for cybersecurity (§\ref{ssec:general}); then, we elucidate the problems of ML solutions developed \textit{in-house} (§\ref{ssec:cots}); and we conclude with the problems related to the adoption of \textit{commercial-off-the-shelf} (COTS) ML products (§\ref{ssec:cots}).

We stress that all problems herein described are \textit{intrinsic}: they can be mitigated to some degree, but the current state-of-the-art does not allow to completely resolve them.

\subsection{General Problems of ML in Cybersecurity}
\label{ssec:general}
Machine Learning follows the so-called ``\textit{indipendent, identically distributed random variables}'' (iid) principle~\cite{dundar2007learning}. Such principle states that the data analyzed during the development of the ML model will be similar to the `future' data that the ML model will analyze after its deployment. If the iid assumption is not met, then the deployed ML model will exhibit a different performance than the expected one (measured during development).
Such iid principle impairs ML deployment in cybersecurity because it interferes with three characteristics of this domain: the \textit{concept drift}, the \textit{adversarial setting}, and the \textit{data confidentiality}. Let us elaborate each of them.

\begin{figure}[!htbp]
    \centering
    \includegraphics[width=1\columnwidth]{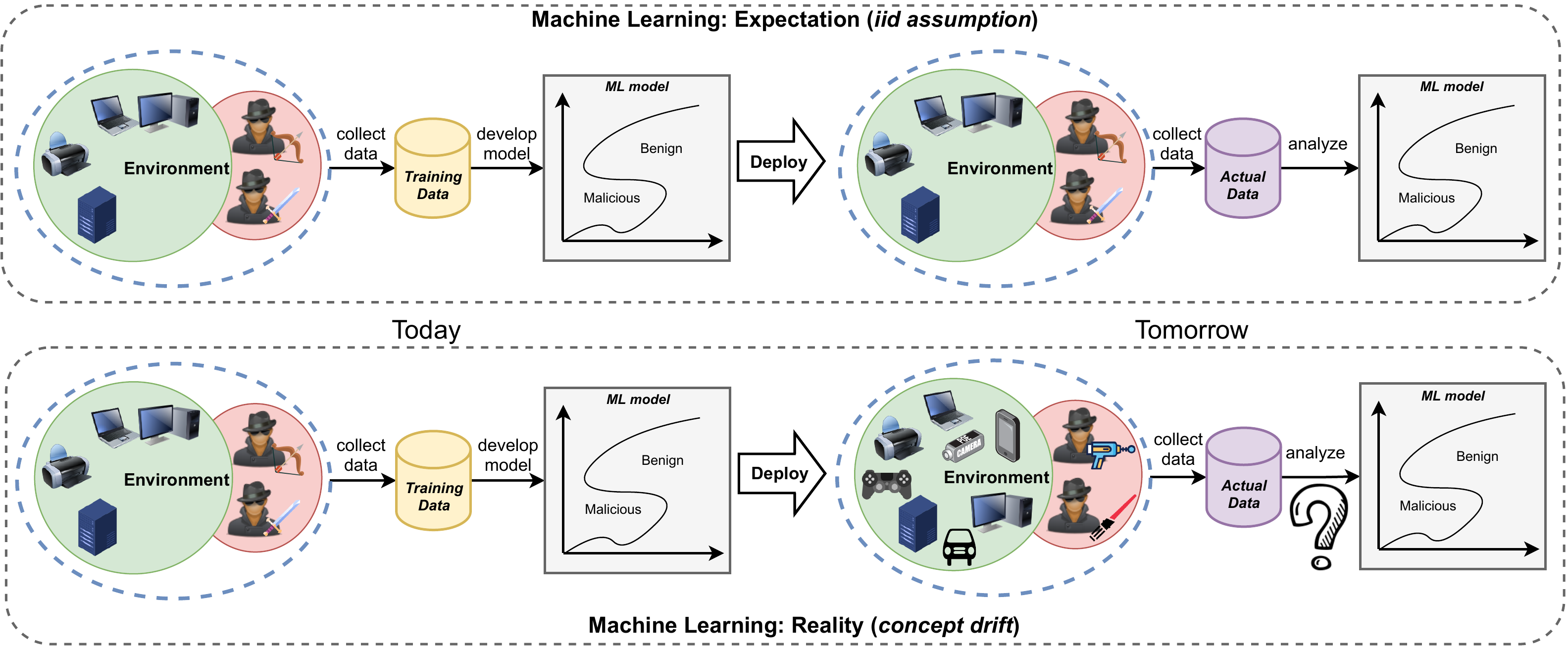}
    \caption{Machine Leaning in the presence of Concept Drift. The ML model expects that the data will not deviate from the one seen during its training. In cybersecurity, however, the environment evolves, and adversaries also become more powerful.}
    \label{fig:drift}
\end{figure}

\paragraph{Concept Drift}
Modern systems are continuously evolving: new devices, services, and even users, are added (or removed) every day. All such mutations contrast the iid assumption, preventing the reliable application of ML \textit{in the long term} because the training data quickly becomes obsolete. This problem is often referred to as \textit{concept drift}, and while it can affect any application of ML, some domains are less touched by it. For instance, in computer vision ``a cat will always be a cat'', allowing to use a ML model trained on the same data for decades---e.g., the ImageNet dataset (collected in 2011) is still used today~\cite{ramanathan2021predet}. This is not the case in cybersecurity, and especially for threat detection: the \textit{environment} constantly changes, and the \textit{adversaries} also adapt. A schematic representation of the concept drift is shown in Fig.~\ref{fig:drift}.

For example, a new vulnerability may be discovered, meaning that some samples previously considered as benign should be treated as malicious\footnote{For instance, hundreds of apps in well-known marketplaces were recently found to be malicious~\cite{kotzias2021did}.}; a new segment may be attached to a network, with a considerably different behavior than the other segments, hence generating a lot of (false) anomalies; finally, attackers can devise novel strategies that cannot be detected by existing mechanisms (e.g., zero-day exploits). As a matter of fact, many research efforts highlighted the significant performance degradation of ML detectors in the presence of concept drift~\cite{jordaney2017transcend, andresini2021insomnia}.
The only practical remedy to concept drift is through constant update of ML systems with new data (labelled, if supervised ML is used) that reflects the current trends.

\begin{figure}[!htbp]
    \centering
    \includegraphics[width=0.7\columnwidth]{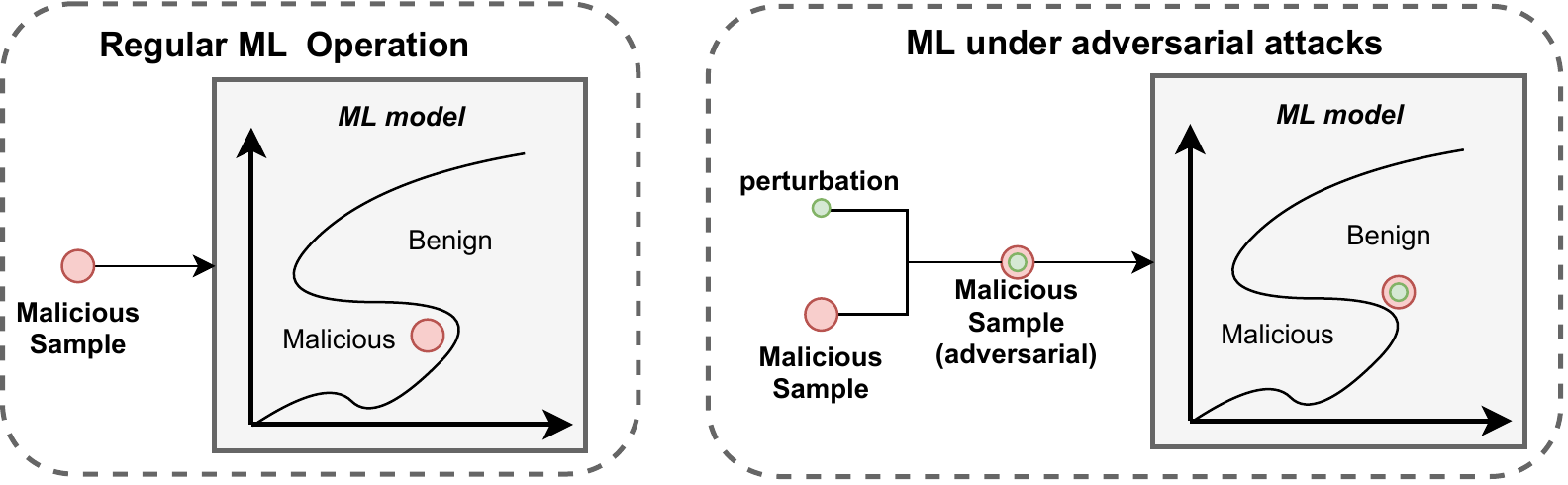}
    \caption{Typical Adversarial Attack against a deployed ML model. By inserting tiny perturbations in the input data, it is possible to fool a ML model and induce an incorrect prediction.}
    \label{fig:adversarial}
\end{figure}

\paragraph{Adversarial Setting}
The cybersecurity domain implicitly assumes the presence of adversaries. Although most attacks are `stationary' (which explains why signature-based methods are still widely employed), motivated adversaries constantly refine and change their offensive strategies. Aside from the risk of zero-day attacks, deployment of ML also exposes to the threat of \textit{adversarial samples}~\cite{laskov2014practical}, which specifically target ML systems. Such threat, schematically depicted in Fig.~\ref{fig:adversarial}, involves applying tiny `perturbations' to some input data with the goal of compromising the predictions of a ML model. Even imperceptible modifications can affect proficient cybersecurity ML detectors. For instance,~\cite{apruzzese2019evaluating} evaded 20 ML botnet detectors by appending a few bytes of junk data to some network communications; whereas~\cite{pierazzi2020intriguing} and~\cite{laskov2014practical} showed a similar effect against ML malware detectors. Even commercial products are affected, such as Google Chrome's phishing detector~\cite{liang2016cracking}. There exist a wide array of strategies to carry out attacks based on adversarial samples, which can affect either the pre- or post-deployment phase of a ML model~\cite{laskov2014practical, apruzzese2019addressing}. 
Despite the proposal of many countermeasures against adversarial samples, (e.g.,~\cite{apruzzese2020hardening, grosse2017adversarial}), no universal solution has been found so-far, and some mechanisms can even decrease the baseline performance (as shown in~\cite{apruzzese2020deep, demontis2017yes}). The best defense, according to Biggio and Roli~\cite{biggio2018wild}, is a \textit{proactive} approach: the adversary must be anticipated and evaluated (and, possibly, countered) \textit{before} ML deployment.

To further stress the importance of such threat, let us clear two \textbf{misconceptions}:
\begin{itemize}
    \item it is common to refer to adversarial samples as `illegitimate'. Such notation is wrong from a security standpoint: any sample (adversarial or not) analyzed by a ML model is considered as \textit{legitimate} (i.e., trusted) by the underlying system that forwarded such sample to the ML model. What is illegitimate is the \textit{attack}, i.e., the application of a perturbation that is specifically crafted to thwart a ML model---but not the adversarial sample.\footnote{To provide a concrete example, let us consider~\cite{apruzzese2018evading}: it is legitimate to increase the size of network communications, but it is illegitimate to do so with the intent of thwarting a ML model. However, a ML model considers all analyzed samples as trusted, because the ML model is oblivious of the intent of the data generation process.}
    \item in related literature, it is common to search for the `minimal' perturbation that allows a sample to thwart a target ML model. However, real attackers are not subject to such constraint.\footnote{For instance, in~\cite{apruzzese2018evading} adding 1KB of data is more effective than adding only 1B. Hence, a real attacker is more likely to add 1KB than just 1B.}
\end{itemize}
The latter observation is crucial for demystifying the effectiveness of the so-called \textit{certified defenses}~\cite{raghunathan2018certified}, which only work if the perturbation is minimal or restricted within a very small boundary.

\paragraph{Confidentiality}
The cybersecurity domain is characterized by its sensitivity to data-privacy, representing a strong barrier for long-term reliance on ML. Let us provide a few examples.
The increasing usage of \textit{encryption} can make some ML systems simply unusable. For instance, a ML-NIDS that inspects the payload of HTTP traffic will not work if the traffic is encrypted via HTTPS---and HTTPS is increasingly replacing the insecure HTTP protocol worldwide. Such problem can also affect other use-cases of ML, such as phishing email detectors: if the emails are encrypted (e.g., via PGP) then it is impossible to analyze their contents with ML. 
Another problematic scenario can involve the analysis of confidential data: the constant changes in data regulation (e.g., the GDPR~\cite{voigt2017eu}) make it difficult to identify data that can be reliably used in the long-term. For instance, consider the approach in~\cite{yen2014epidemiological} (cf. §\ref{ssec:risk}), which leverages (among others) user information to estimate the infection risk. Such approach could not be applied today without the explicit consent of all the users of a company. 
Moreover, both of these issues (confidential and encrypted data) also impair \textit{labelling} procedures, because it is not possible to (manually) verify the ground truth of a sample if such sample cannot be `seen' by a human expert.
Finally, it is understandable that enterprises do not want to publicly disclose their data, generating an overall shortage of publicly available datasets that can be used to evaluate ML systems~\cite{sharafaldin2018toward}. Although this latter problem primarily affects \textit{research}, it also implicitly affects \textit{practice} because showing a ML system that works in different settings can foster its adoption in real scenarios. We discuss potential solutions to the limited data availability in §\ref{ssec:data}.

\subsection{Problems with in-house development of ML systems}
\label{ssec:inhouse}
Despite the problems presented in §\ref{ssec:general}, an organization may be willing to create a completely in-house ML solution. In this case, the organization can fully control the scope, data and overall suitability of the resulting ML model. However, such advantage comes at a price: the ML model must be first \textit{developed} and must also be \textit{maintained}. Both of these procedures are challenging in cybersecurity.

\paragraph{Initial Development}
Developing the initial ML model requires three steps:
(i) selecting a ML algorithm; (ii) finding the right data; and (iii) fine-tuning the performance. In some domains, these steps are almost straightforward. For instance, in computer vision it is established that deep learning algorithms outperform others; moreover, suitable data (potentially labelled) is easier to acquire---either because it is publicly available (e.g., ImageNet~\cite{ramanathan2021predet}) or because it can be cheaply produced (e.g., the popular captchas~\cite{bursztein2011text}). Unfortunately, none of these advantages apply to cybersecurity. For instance, some research works show that deep learning is worse (e.g.,~\cite{apruzzese2020deep, pontes2021new}) while others claim the opposite (e.g.~\cite{nugraha2020performance, vinayakumar2019deep}). Similarly, there is confusion with respect to which \textit{features} should be taken into account (cf. §\ref{ssec:phishing} where~\cite{basnet2014learning} use 130 features and~\cite{sahingoz2019machine} use 30, achieving similar performance). 
Finding the right data is also inherently more challenging in cybersecurity. Such challenge includes acquiring data of \textit{high quality} and in the \textit{right amount}.
Labelling requires expert knowledge, and according to~\cite{miller2016reviewer} a company cannot afford to label more than 80 malware samples \textit{per day}. For reference, the initial deployment of Tesseract~\cite{pendlebury2019tesseract} required 50 thousand labelled samples. Unlabelled data may be easier to acquire, but as shown in §\ref{sec:beyond} it can come from heterogeneous sources and be in different formats, requiring a detailed preprocessing pipeline to collect, store, and forward such data to the ML model. Furthermore, the iid assumption (cf. §\ref{ssec:general}) prevents a reliable use of data originating from different environments~\cite{sommer2010outside}, hence even the (few) publicly available data can have questionable effectiveness. 
Finally, a common \textbf{misconception} is thinking that the performance of a ML model is linearly dependant on the size of its training data\footnote{According to the founder of Deep Learning, Andrew Ng, this is also becoming true for Deep Neural Networks~\cite{andrew2021unbiggen}.}: in some cases, \textit{smaller datasets can yield to superior} ML models---we will show this in our case studies (§\ref{app:montimage}). Nevertheless, any given dataset must also be \textit{balanced}: in real environments, a malicious event is a rare occurrence and a given dataset should reflect such distribution~\cite{wheelus2018tackling}. 

To aggravate all of the above, it is not possible to determine \textit{a priori} which combination (algorithm, features, dataset, balancing) yields the best performance after deployment. Hence, empirical and time consuming evaluations---by training and testing multiple ML models---are always a necessity. As a result, finding the most optimal tuning for real deployments may require a huge amount of manual effort by trial-and-error.

\paragraph{Constant Maintenance}
To mitigate the disruptive effects of concept drift (§\ref{ssec:general}), it is fundamental to continuously update a given ML solution with data reflecting the current trends. Such procedures are costly but can be alleviated via lifelong learning solutions (cf. §\ref{ssec:raw}). However, a common \textbf{misconception} is that `update' procedures simply entail finding (and, if necessary, labelling) new data. This is an underestimation, because such procedures also require to: (i) decide what to do with `old' data; and (ii) finding the `sweet spot' that yields the adequate performance.
Indeed, maintaining old data can be detrimental in some cases (e.g., if some `benign' samples are discovered to be `malicious'), but completely removing it can also adversely affect the performance (e.g., some `old' phenomena can reappear in the future). Nonetheless, even small changes in the training data can decrease the performance of a ML system (e.g., this is the fundamental principle of poisoning attacks~\cite{apruzzese2019addressing}). These issues require additional manual labour through trial-and-error.

A potential mitigation for all such tuning operations (both pre- and post-deployment) may come in the development of techniques focused on \textit{explaining} the decisions of ML systems (e.g.~\cite{marino2018adversarial}), which are currently difficult to interpret---especially for deep learning~\cite{amarasinghe2018toward}. This is an intriguing direction of research, which has very recently also touched the area of adversarial ML (e.g.~\cite{amich2021explanation, demontis2019adversarial})
\subsection{Problems of Commercial-off-the-Shelf ML products}
\label{ssec:cots}
Developing an in-house ML model may be prohibitive (e.g., in terms of computational or human resources), and COTS solutions represent a viable alternative. In this case, all the operations presented in §\ref{ssec:inhouse} must be performed by the product vendor. However, we point out two drawbacks of such COTS solutions, aggravated in cybersecurity scenarios. Specifically, such solutions implicitly have a \textit{limited scope}, and they may (inadvertently) suffer from lack of \textit{transparency}.

\paragraph{Limited Scope}
Relying on third-party solutions limits any end-user to their intended scope, meaning that some tasks simply cannot be accomplished with products currently on the market.
For instance, any commercial ML model cannot be trained on the exact data used by an organization---at least initially. The organization can allow the vendor to collect their data, and use such data to refine the ML model; however, this may not be possible due to confidentiality reasons (§\ref{ssec:general}). Therefore, some commercial solutions can be used only if the deployment environment (of the organization) resembles the pre-deployment environment (of the vendor) used to generate the data for the corresponding ML model. For example, phishing websites are malicious `everywhere', meaning that it is possible to \textit{transfer}~\cite{yin2020apply} ML phishing detectors. However, such transferring cannot be easily done for other cybersecurity tasks, such as NIDS~\cite{apruzzese2022cross}. This is because \textit{every network is unique}~\cite{sommer2010outside}, and a malicious behavior in one network can be benign in a different network. Due to such issue, most COTS products leverage (unsupervised) ML, and mostly for anomaly detection (e.g.,~\cite{Lastline:AI, Darktrace:ML}).

\paragraph{Lack of Transparency}
COTS solutions come as a `black-box', and the decision to deploy such solutions depends on their advertised performance. This fact leads to many issues, all sharing a common culprit: the cost of \textit{misclassifications} in cybersecurity.
In some domains, incorrect predictions do not have severe consequences: for instance, a recommender ML system (e.g., the one in AirBnb~\cite{haldar2019applying}) that makes an incorrect recommendation is not a cause of concern. 
In contrast, in cybersecurity a single false negative can be the difference between a compromised and a secure system. At the same time, both employees and security analysts are annoyed by false alarms, which can even be exploited by attackers to conceal more severe threats~\cite{chivers2013knowing}. 
By considering the performance metrics reported in Table~\ref{tab:metrics} (cf. §\ref{ssec:glossary}), we remark that each metric focuses on a single aspect, and even good scores can be meaningless if not contextualized.\footnote{As an example, consider a detector evaluated on a dataset containing 9990 benign samples and 10 malicious samples: an Accuracy of $99.99\%$ can be obtained by only detecting 1 malicious sample (out of 10), despite its inability to detect $90\%$ of the attacks. Another example is a FPR of $1\%$: it may appear low, but if the environment generates 300k alarms (as in~\cite{nadeem2021alert}) such FPR corresponds to 3000 false alarms. Note that also the inverse is true: an increment of just $1\%$ in the TPR can be either an almost negligible or an extremely significant performance boost.} Nonetheless, even if a COTS ML solution is fully transparent (i.e., all metrics are reported and contextualized), the performance will always refer to the environment of the \textit{vendor}, which is likely to differ from the real deployment setting. Finally, we mention that---to the best of our knowledge---no COTS ML solution (including those not pertaining to security tasks) reports its robustness to potential adversarial attacks, which is a severe deficiency in cybersecurity scenarios.

\takeaway{In cybersecurity, ML can provide great benefits but also presents many risks due to the intrinsic adversarial setting and the dynamic ecosystem. Such risks must be taken into account today, and should be addressed by future works.}
\section{The Future of Machine Learning in Cybersecurity}
\label{sec:future}
We have elucidated the benefits (§\ref{sec:detection} and §\ref{sec:beyond}) as well as the problems (§\ref{sec:issues}) of ML for cybersecurity. There are potentially infinite ways to advance the state-of-the-art, such as increasing existing performance (e.g.,~\cite{dias2020go}),
mitigating known issues (e.g., the poor explainability~\cite{samek2017explainable, amich2021explanation}), as well as development of novel applications of ML in cybersecurity (e.g., the integration of quantum computing~\cite{gouveia2020towards}).

\begin{figure}[!htbp]
    \centering
    \includegraphics[width=0.8\columnwidth]{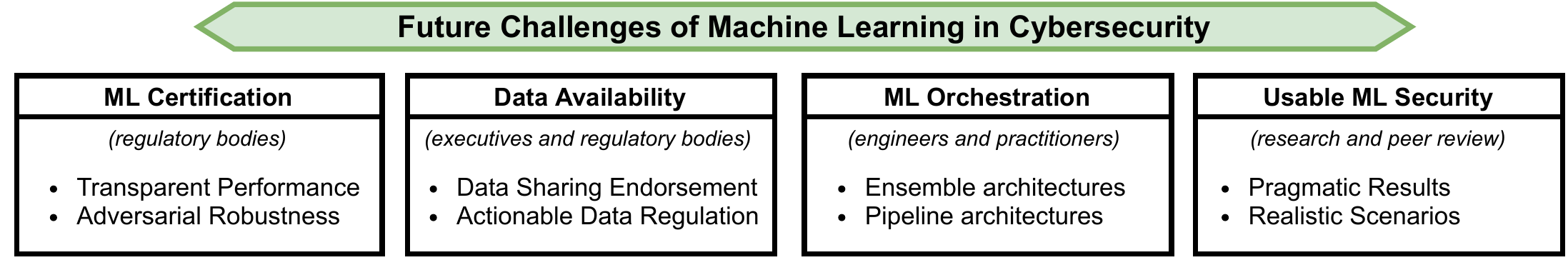}
    \caption{Future Challenges of Machine Learning in Cybersecurity. Addressing all such challenges requires the cooperation of four players: regulatory bodies, corporate executives, engineers, and researchers.}
    \label{fig:future}
\end{figure}

As a constructive step forward, this section highlights which future developments can \textit{completely revamp} the state-of-the-art of ML in cybersecurity. Although every improvement is appreciated, we believe that the existing gap between research and practice can only be closed by the joint contribution of four players: \textit{regulatory bodies}, \textit{corporate executives}, \textit{engineers}, as well as the \textit{research community}. Specifically, we identify four future challenges that---if properly addressed---can revolutionize ML in cybersecurity.
We now elucidate these challenges (schematically shown in Fig.~\ref{fig:future}), explaining their root causes and our recommended course of action for each of the four `players' indicated above.



\subsection{Certification (Sovereign entities)}
\label{ssec:sovereign}
The 2020 EU White Paper on Artificial Intelligence~\cite{Europe:AI}---also followed by a 2021 US DHS report~\cite{DHS2021AI}---indicates \textit{trustworthiness} as one of the key requirements for future ML applications. Especially, emphasis is put on `high-risk' scenarios, where deployment of ML should conform with pertinent legal requirements. Cybersecurity applications naturally qualify as high-risk, hence procedures that certify the \textit{performance} and \textit{robustness} of ML systems should be developed and enforced by regulatory bodies. Let us elaborate.

\paragraph{Performance Certification}
Comprehensive testing represents the only instrument for performance verification of a ML system. However, despite hundreds of works, there is a \textbf{lack of standardized evaluation protocols}. This is a problem especially for COTS products, as performance assessments may be carried out in biased environments, or may consider unfair comparisons that inflate the results to favor a given ML solution. Meaningful assessments must consider the realistic distribution of data and take into account the (likely) temporal shift. Traditional cross-validation techniques, typical for ML in the computer vision domain, should be used only for tuning: specifically, the performance should be validated via statistical tests. Establishing standardized evaluation protocols would foster pragmatic and fair comparisons, promoting overall ML deployment in practice. Nevertheless, the full details of such operations (e.g., the data used, the evaluation methodology, and the final results) should be transparent to the customers of COTS ML systems. 

\paragraph{Robustness Certification}
The increased interest towards ML led to (scientific) investigations of its robustness in adversarial scenarios, bringing to light the vulnerability to adversarial examples (§\ref{ssec:general}). Yet, no universal solution has been found so far, with some defenses being broken in the time span between their appearance as a preprint and their publication as a peer-reviewed article\footnote{For instance, defensive distillation was proposed in 2016~\cite{papernot2016distillation} and broken few months later~\cite{carlini2016defensive}.}. The first step to solve this problem is to \textbf{acknowledge that no ML solution is flawless}. Indeed, to quote a recent survey on the cybersecurity perspective of European stakeholders~\cite{fischer2021stakeholder}: ``\textit{security of ML and adversarial attacks was not mentioned as one of the key challenges by the interviewees.}'', which epitomizes that such threat is not perceived by the end-users of ML solutions.
To address these issues, assessments of adversarial robustness must become mandatory in evaluations of \textit{any} ML-based solution for cybersecurity. The most likely security risks, and their potential consequences, should be known \textit{before} real ML deployments. Moreover, all the details of such assessments should be transparently provided.

\recommendation{To ensure better transparency and reliability, regulatory bodies must enforce the development and adoption of standardized procedures that certify the performance and robustness of ML systems.}

\subsection{Data Availability (executives and legislation authorities)}
\label{ssec:data}
The effectiveness of any ML-solution depends on the data used to train the corresponding ML model. However, among the toughest challenges faced by ML in cybersecurity is \textit{finding appropriate data}. Despite the recent interest in ML led to the release of more open datasets (e.g.,~\cite{ring2019survey, arp2014drebin, allix2016androzoo}), such datasets exhibit limitations~\cite{verma2019data}. For instance, inaccurate labels, fast obsolescence, small and synthetic environments (e.g.,~\cite{moustafa2015unsw}), or even flawed generation process---as shown in~\cite{engelen2021troubleshooting}. All these problems can only be mitigated to some degree (e.g.~\cite{bertoli2021bridging}), and cannot be solved by the scientific community.
The lack of adequate data (i.e., \textit{real} and \textit{labelled}) makes evaluations of ML conducted in research environments to be of questionable value, preventing sound assessments of ML capabilities, and ultimately hindering its deployment.
Addressing the shortage of data is possible, but it requires the joint intervention of industrial stakeholders and regulation authorities: the former should promote \textit{data sharing}, the latter should devise more \textit{actionable data regulation} policies.

\paragraph{Data Sharing}
A solution to the lack of adequate data is the promotion of data sharing practices. In cybersecurity, some portions of data can be easily shared: for instance, Sophos has recently released over 20M (labelled) malware samples~\cite{harang2020sorel20m}; similarly, the recent CrimeBB dataset~\cite{pastrana2018crimebb} contains 1 million accounts crawled from darkweb forums for 10 years. In contrast, other pieces of data (especially \textit{benign} data) are more confidential and hence their \textbf{disclosure requires explicit permission from corporate executives}. Acquiring such permission is a tough barrier, especially due to privacy and secrecy issues. However, we observe that sensitive information can be anonymised (e.g.,~\cite{potiguara2020big}), and recent advances in \textit{federated learning} overcame such problems~\cite{dayan2021federated}. 

There indeed exist some success stories of data sharing \textit{platforms} focused on security information, such as the EU-OF2CEN project~\cite{spagnolettia2020digital}. Similar platforms represent a great \textit{opportunity} for some companies, as they open the doors to a new market entirely dedicated to ML datasets, potentially with (updated) ground truth (e.g.,~\cite{song2021try}).
From this perspective, a promising initiative is STIX CyBox~\cite{sadique2018automated}: its goal is creating a threat intelligence platform shared by multiple parties, facilitating the entire process of incident detection and response. Nonetheless, such platforms must (i)~contain unbiased data---otherwise there a the risk of manipulating future developments~\cite{maschmeyer2021tale}---and (ii)~comply with the existing regulation, hence requiring the involvement of the respective authorities.

\paragraph{Actionable Data Regulations}
The strategical importance of data gave birth to multiple regulations that `protect' data owners and limit abuse of sensitive information. Despite ensuring more privacy rights, such regulations introduced additional constraints on data gathering and processing, resulting in yet another barrier to ML developments---both for research and practice.
Specifically, the (already costly) \textbf{data-labelling procedures are crucially affected by such regulations} (§\ref{ssec:general}). Even if action is taken by executives to disclose their corporate data, existing regulation policies are difficult to interpret and likely to change in the future~\cite{nweke2020legal}: for instance, information that can be shared `today' may not be shareable `tomorrow', hindering long-term projects. However, we observe that some GDPR compliant data-sharing platforms exist (e.g.,~\cite{horak2019gdpr}). Hence, the regulatory authorities should promote such efforts even in the cybersecurity context, for instance by providing actionable policies that ensure the compliance of (open) data in the long-term.

\recommendation{To address the shortage of adequate data, companies should be more willing to share data originating in their environments (e.g.,~\cite{spagnolettia2020digital}), whereas regulation authorities should promote such disclosure by defining proper policies and incentives~\cite{nweke2020legal}.}
\subsection{Usable Security Research (scientific community)}
\label{ssec:research}

The combination of ML and cybersecurity is a fertile opportunity for research, and recently inspired many related papers. Such trend, however, is a double-edged sword. On the good side, the rising scientific interest demonstrates the high potential of ML for cybersecurity. On the bad side, such abundance can be \textit{detrimental} for real ML deployments, as it may raise more questions rather than provide answers. Specifically, we identify two problems of existing research: the \textbf{lack of pragmatic results}, and the \textbf{limited consideration of realistic scenarios}.

\paragraph{Pragmatic Results}
One of the primary goals of research is to ``outperform the state-of-the-art''. In the context of ML, such goal requires proposing a novel ML method, and then show that such method achieves a better performance than prior works---an objective that can be achieved without providing any `true' contribution to the state-of-the-art. For example, by slightly changing the training data it is possible to achieve a superior performance; similarly, an existing solution may be sub-optimally reproduced (by using, e.g., a different dataset, or different tuning parameters). Note that all such `flaws' can be \textit{unconsciously} introduced by researchers\footnote{A recent paper describing the pitfalls of ML in cybersecurity is~\cite{arp2020dos}}. This phenomenon, also referred to as \textit{benchmark lottery}~\cite{dehghani2021benchmark}, results in an overall confusion on what really works best and impairs real ML deployments. Among the main culprits of such phenomenon is the \textbf{poor reproducibility} of researches, as very few works disclose the entire information required to replicate their experiments. Therefore, novel researches cannot properly reproduce previous works; and the peer-review process cannot assess whether the experimental protocol is correct and unbiased. At the same time, however, we point out that most scientific venues do not allow (or require) inclusion of any supplementary and technical resource. Hence, even researchers must face a difficult decision about what low-level information should be included in the actual submission---which is subject to page limitations.
\recommendation{The peer-review process should facilitate and enforce the inclusion of the material for replicating ML experiments. At the same time, such material should be evaluated to ensure its correctness---potentially by a separate set of reviewers with more technical expertise.}

\paragraph{Realistic Security Scenarios}
As a direct consequence of the benchmark lottery phenomenon, many research papers simply focus on providing `better numbers' than past work, overlooking the assumptions made by such past work. In the context of cybersecurity, this is a problem because realistic circumstances must be considered, and any result that stems from unrealistic scenarios is of questionable value.
For instance, there is a \textbf{superficial treatment of training data}: only few papers (e.g.,~\cite{andresini2021insomnia, pendlebury2019tesseract}) consider the \textit{concept drift}, which is intrinsic in cybersecurity; moreover, many recent papers (e.g.,~\cite{su2020bat}) still use \textit{outdated datasets}, such as the NSL-KDD which is over 20 years old and does not reflect any current environment. The result is that all papers propose ML methods that achieve near-perfect performance---but what is the practical impact of all such researches? We acknowledge that public (labelled) data is difficult to acquire, but in the last years several datasets have been openly released (e.g.,~\cite{sharafaldin2018toward, moustafa2015unsw}).
The impression is that the cybersecurity setting is turning into a yet-another research playground where new ML methods are evaluated on some `security-related' data, but realistic security considerations are only made in the introduction to provide some justification for a given publication venue. 
Specifically, there is a \textbf{lack of realistic threat models}. Such lack is epitomized in the emerging field of adversarial ML (§\ref{ssec:general}), where most attacks against security systems assume extremely powerful opponents. For instance, the authors of~\cite{apruzzese2021modeling} show that the majority of attacks against ML-NIDS require adversaries with direct access to the ML-NIDS itself, which is an assumption that violates the basic security principles. Similarly,~\cite{pierazzi2020intriguing} show that adversarial attacks have a different effectiveness when the opponent cannot manipulate the data-processing pipeline (which is usually not accessible). Hence, it is not surprising that the industrial stakeholders are either confused or do not care about adversarial examples---as evidenced by two recent surveys~\cite{kumar2020adversarial, fischer2021stakeholder} and the detailed case study in~\cite{boenisch2021never}.
\recommendation{Future researches on ML applications for cybersecurity should have a closer connection with the real world. The assumed \textit{threat model} should be realistic, the \textit{dataset} should resemble recent trends, and the \textit{concept drift} should be taken into account.}
\subsection{Orchestration of Machine Learning (engineers)}
\label{ssec:engineers}
ML is not meant to fully replace existing systems or human experts. Rather, it should provide an additional `perspective' that can be used to identify otherwise overlooked phenomena. However, ML methods exhibit huge variance (e.g., different performance~\cite{verma2019phishing, pontes2021new}, or adversarial robustness~\cite{apruzzese2019evaluating}), and a single ML solution cannot protect against all threats that can target modern organizations.\footnote{The most exemplary use-case are zero-day attacks, which can easily evade supervised ML methods: zero-day samples cannot---by definition---be included in the training data. Anomaly detection through unsupervised ML is more feasible, but at the cost of many false positives.}
Addressing all such issues is possible by \textit{orchestrating} diverse ML solutions. Indeed, any ML model (irrespective of its goal) ultimately represents just a single component of a cybersecurity system---which can be a `hybrid' system that leverages also non-ML techniques. However, such orchestration requires the expertise of \textit{ML engineers}, who must coordinate different outputs to extract actionable information. Specifically, ML (and non-ML) models can be combined either in an \textit{ensemble} or in a \textit{pipeline} architecture, depending on the final goal of the system. 

\paragraph{Ensemble architecture}
One of the most proficient ways to combine different ML models is the so-called \textit{ensemble}~\cite{biggio2015one}. The idea is leveraging many simplified learners \textit{with a common goal}: each ML model of the ensemble analyzes the same data, but by focusing on a specific problem. For instance, it is possible to create ML-NIDS using ensembles of ML models, in which each model has the same goal (i.e., intrusion detection), but focuses on a specific threat (e.g., botnet or DoS attacks~\cite{mirsky2018kitsune}). Despite the proven performance benefits of such architectures, a tough challenge faced by engineers is the \textbf{lack of standardized feature sets} that can be used to devise all such systems. Each model of the ensemble must ultimately analyze the same data, and depending on the features provided as input the performance can greatly differ (as shown in~\cite{binbusayyis2019identifying}). Our industrial case studies in §\ref{sec:casestudies} consider a similar architecture.

\paragraph{Pipeline architecture}
When the system envisions ML models having systematically \textit{different inputs and outputs}, they must be organized in a pipeline architecture.
For example, it is possible to create an ensemble of ML models for threat detection (§\ref{sec:detection}), and then use their outputs for threat intelligence (§\ref{sec:beyond}).
Similar systems already exist, either as COTS products (e.g., SIEM\footnote{System Information and Event Managers: \url{https://www.forcepoint.com/cyber-edu/siem}} or SOARS \footnote{Security Orchestration Automation and Response Systems: \url{https://www.rapid7.com/solutions/security-orchestration-and-automation/}}) or as scientific proposals: for instance, ARCUS~\cite{yagemann2021arcus} is a security-focused orchestration platform that could benefit from the integration of many of the ML solutions discussed in this paper. However, such architectures are challenging to implement by engineers: \textbf{each individual component is affected by all the issues} presented in §\ref{sec:issues}, therefore multiplying their impact.

\recommendation{Orchestrating complex systems that use (combinations of) ML and non-ML solutions is beneficial for cybersecurity. Hence, ML engineers and practitioners should clearly highlight how to combine all such components in order to maximize their practical effectiveness.}

\section{Case Studies: Industrial Applications of ML for Cybersecurity}
\label{sec:casestudies}

As a final contribution of this paper, we present two case studies that showcase real and successful industrial applications of ML in cybersecurity. Many commercial products are advertised as leveraging ML. Yet, most of these products are provided as black-boxes, preventing any understanding of how ML is actually applied \textit{in practice}. Specifically, our case studies involve the two following scenarios:
\begin{itemize}
\item  using ML for detecting Cache Poisoning Attacks against Named Data Networks. The approach is integrated in a NIDS developed by \montimage{} (§\ref{app:montimage});

\item combining sequential deep learning with non-ML methods for protecting Industrial Control Systems. The approach is integrated in a cybersecurity device developed by \sgrupo{} (§\ref{app:s2grupo}).
\end{itemize}
Both of these solutions use ML for anomaly detection with limited supervision.
Our goal is to provide a high-level overview on such `black-box' ML systems by elucidating their internal functionalities.\footnote{The commercial nature of such systems---which are built on the end-users data---makes some low-level details to be protected by NDA, but explicit requests can be made by contacting the respective vendors.}.

\subsection{Detection of Cache Poisoning Attacks in Named Data Networks}
\label{app:montimage}

Information Centric Networking (ICN) is a revolutionary paradigm in the context of communications: while most of the Internet follows a host-to-host perspective, ICN adopts a host-to-content vision~\cite{AhlgrenDIKO12}. The ICN architecture is more suitable for massive content diffusion (e.g., video streaming), representing the major use cases of modern networks. Despite providing multiple benefits in terms of bandwidth efficiency and scalability, ICN can fall victim to Denial of Service (DoS) attacks and, in particular, to poisoning attacks~\cite{xu2015elda, afanasyev2013interest}. In this case study, we analyze a real ML detection system that protects against such attacks targeting ICN architectures. The specific techniques are integrated into the Montimage Monitoring Tool\footnote{\url{https://montimage.com/products/MMT\_DPI.html}} (\mmt{}), which is a module of the IDS framework developed by \montimage{}~\cite{nguyen2018security, MMTarticle}.

\paragraph{Scenario and Challenges}
This case study focuses on the well-known ICN approach of Named Data Networking (NDN)~\cite{zhang2014named}. Such NDN approach leverages a pull-based mechanism using two kinds of packets: \textit{Interest} (a request for a content) and \textit{Data} (the response with the content).
When a given user wants to retrieve some content, the user (i) specifies the desired content's name (e.g. ``/data/video.mp4'') in an Interest, (ii) sends such Interest through the NDN network, and (iii) receives the corresponding Data---which can be provided either by the content \textit{producer}, or by any intermediate NDN node storing a copy of such Data. The practical implementation of NDN exposes to the risk of new security attacks, such as the Content Poisoning Attack (CPA)~\cite{xu2015elda}.
In CPA, a malicious \textit{producer} (content creator) colludes with a malicious \textit{consumer} (a user requesting content) to force any NDN node on their path to insert malicious content in their content storage (CS), hence causing poisoning attacks. This results in nodes answering some requests with such malicious content: for example, a victim may ask for a specific webpage and instead be redirected to a malicious phishing website. CPA are a dangerous threat to NDN, as shown in~\cite{nguyen2017content}: analyses on real system highlighted that identifying CPA is \textit{impossible} via static and human-based approaches. This is due to the intrinsic characteristics of NDN, as each node in the network topology reacts differently. Moreover, NDN are also susceptible to Interest Flooding Attacks (IFA), which represent a variant of DoS in which the NDN is `flooded' with interest requests~\cite{signorello2017advanced} for existing or even non-existing content that can disrupt the distribution of content. Although IFA are easier to identify than CPA, countering \textit{both} IFA and CPA is challenging and requires the usage of more dynamic analytical techniques---such as ML.

\paragraph{\montimage{} ML-Solution} 
The ML-solution developed by \montimage{} leverages \textit{ensembles} of ML models organized in a Bayesian Network Classifier (BNC)~\cite{NguyenMDCC17}. The intuition is that detection of CPA is only possible by monitoring the behaviour of each node in a NDN network---and, specifically, by analyzing and cross-correlating the evolution of different metrics for each node.

Such goal is achieved by means of specific probes deployed on each node and monitoring its complete activity. In particular, each probe collect metrics related to the Data plane of NDN: Content Store (CS), Pending Interest Table (PIT), Faces. The latter, in particular, are an abstraction of a communication channel that NDN uses for packet forwarding. Such abstraction represents data coming from diverse `faces', i.e.,: overlay tunnels over TCP and UDP; delivery of NDN network layer packets (e.g., Interest, Data packets); inter-node communication channels that send packets to other nodes; and intra-node communication channels that send packets to another process on the same node.

The information captured by these probes is then analyzed by ensembles of \textit{micro-anomaly-detectors}, each focusing on deviations from the normal behaviour of a single metric captured by each probe. It is true that CPA can impact many metrics and in different ways, raising hundreds of (likely) false alarms by each micro-detector. However, \textit{correlating} all the alarms with a BNC allows to (i) increase the detection performance while (ii) mitigating the high rate of false alarms generated by individual micro-detectors.

\begin{figure*}[!htbp]
	\centering
	\includegraphics[width=0.9\columnwidth]{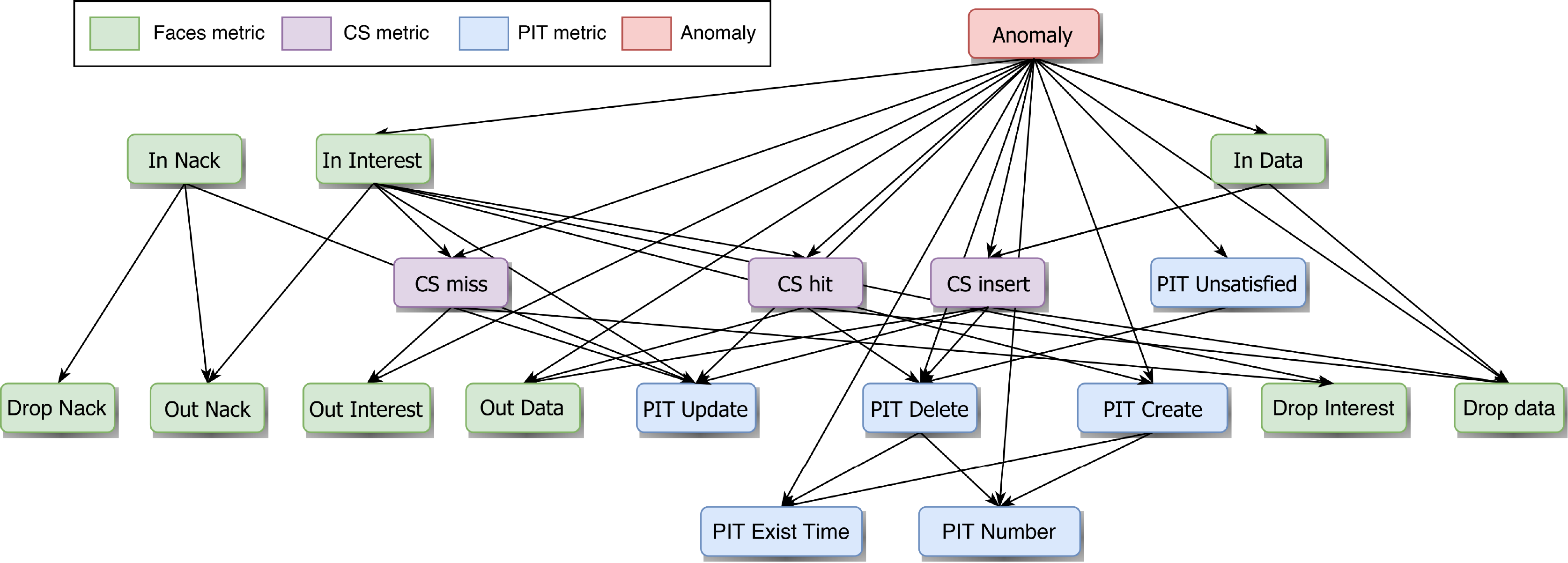}
	\caption{Architecture of the Bayesian Network Classifier adopted by \montimage{} to detect CPA in NDN. Each node represents a micro detector that focuses on a single metric. The Anomaly node correlates the output of all other NDN nodes.}
	\label{fig:bayesnet}
\end{figure*}

A schematic representation of the considered BNC is provided in Fig.~\ref{fig:bayesnet}: the `anomaly' node (denoted in red) represents the anomalies that can occur in the entire NDN, whereas the remaining nodes represent the individual micro-detectors. Hence, each node focuses on a single metric, specifically: Faces, CS, or PIT (denoted in green, purple and blue in Fig.~\ref{fig:bayesnet}).
The (directed) edges in the BNC represent the causal relationships between the Anomaly node and a metric (or pairs of metrics).
An edge connects the `causing' node to the `affected' node.
The causal relationships are deduced based on the processing of each packet arriving to the NDN node.

\begin{wrapfigure}{r}{0.40\textwidth}
  \begin{center}
    {\includegraphics[width=0.35\columnwidth]{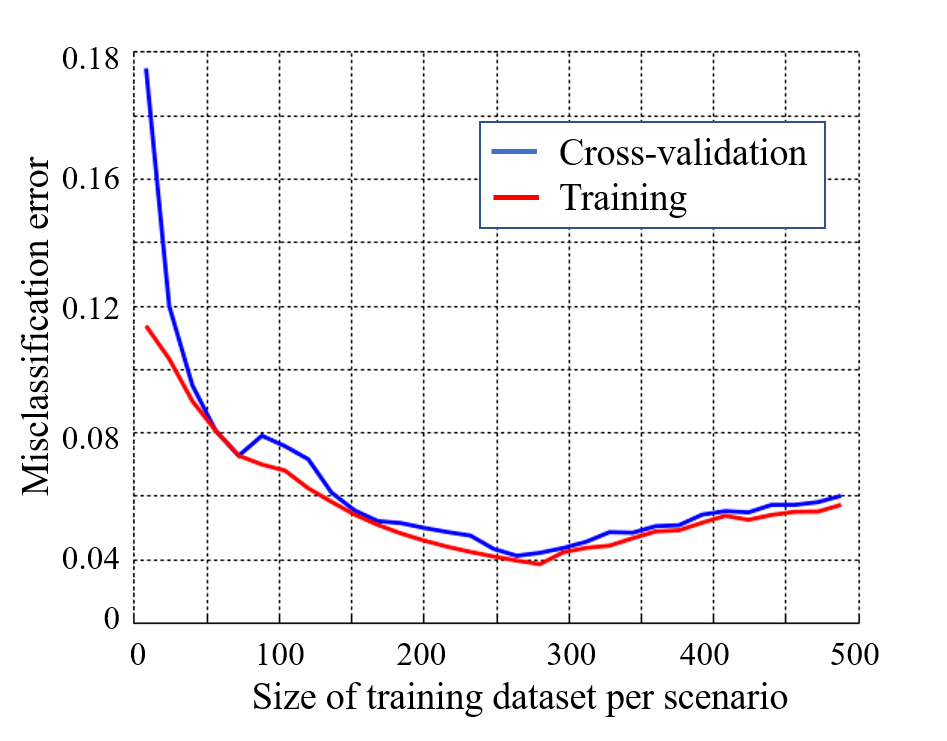}}
  \end{center}
  \caption{Preliminary assessment of the BNC to identify the optimal size of the training dataset.}
  \label{fig:train_test_error}
\end{wrapfigure}

\paragraph{Evaluation and Results}
It is necessary to conduct a preliminary assessment of the learning efficency of the BNC before its deployment. This is because NDN generate a lot of traffic, and even though the BNC can `condense' the raised alarms it is still important that such alarms---and, specifically, false alarms---are within acceptable levels.
To this purpose, \montimage{} first collects huge amounts of \textit{real} data from the probes, and then uses such data (assumed to be benign) to train (and test) a BNC. Specifically, multiple BNC are assessed, each considering a different training size: the goal is finding the optimal size that minimizes the rate of false alarms. The results of such assessment are reported in Fig.~\ref{fig:train_test_error}, showing the misclassification error (as measured via 5-fold cross-validation) as a function of the training size.
We observe that an optimal value is achieved when the training set contains $\sim \! 280$ samples\footnote{We observe that such samples represent \textit{alarms} corresponding to multiple signals, and not to raw events.}. For higher values, the error increases due to overfitting (this phenomenon confirms the misconception outlined in §\ref{ssec:inhouse}). Thus, for the considered deployment scenario, \montimage{} uses training sets of 280 samples---corresponding to 23 minutes of \textit{real} reportings.

To evaluate the performance in production settings, \montimage{} reproduces the NDN topology in~\cite{0001ABJcCPWZ14} and creates two distinct environments, each adopting a specific NDN routing strategy: \textit{bestroute} or \textit{multicast}. Then, each environment is monitored for 10 minutes, and the attack is simulated in the last 5 minutes. Specifically, multiple CPA are launched, each considering increasing \textit{payloads}, denoting the number of requests for content (i.e., Interests) per second; in our case, we consider payloads of 5, 10, 20 and 50 Interests per second. In comparison, legitimate clients produce 10 Interests per second (on average): hence, the malicious traffic ranges from half to five times the legitimate traffic. The traffic generated during such simulations is collected and used to assess the quality of the BNC: the goal is verifying whether the BNC is capable of identifying the CPA, which occurs in the last 5 minutes.

To provide a twofold perspective of the performance (see §\ref{ssec:cots}), \montimage{} measures the True and False Positive Rate (TPR and FPR---cf. Table~\ref{tab:metrics} in §\ref{ssec:glossary}). The results of such evaluation, performed on a testing set of 240 samples, are reported in Table~\ref{table:cpa_results}. We observe that the TPR increases for greater payloads, because the CPA become more conspicuous. Nonetheless, it is appreciable that even CPA with low payload can be effectively detected. Finally, the low FPR is crucial for real deployments as they are annoying to human operators. 
All such results are due to the advantages provided by the BNC, because BNC use a probabilistic approach allowing to take into account the underlying random nature of the observed metrics. Such property makes BNC tailored for multi-variate anomaly detection in \textit{real} environments.
In contrast, other ML algorithms present significant drawbacks: for instance, `deep' neural networks are excessively difficult to develop in such settings (also due to their poor explainability); whereas other `shallow' algorithms, such as SVM, simply do not allow to efficiently represent and correlate all the metrics affected by CPA.

\begin{table}[!htbp]
	\begin{center}
		\caption{CPA detection performance for two different routing strategies and increasing attack payload.} 
		\begin{tabular}{ |c||c|c|c|c|c|} 
			\hline
			 \makecell{Routing\\Strategy} &  \makecell{Attack payload \\ (\# Interest/s)} & 5 & 10 & 20 & 50  \\
			\hline
			\hline
			\multirow{2}{*}{\textit{Bestroute}} & \% True Positive & 95.0\% & 95.3\% & 97.0\% & 98.3\% \\				
			\cline{2-6}
			& \% False Positive & 1.0\% & 1.0\% & 1.0\% & 1.0\%	\\	
			\hline		
			\multirow{2}{*}{\textit{Multicast}} & \% True Positive & 63.3\% & 72.8\% & 79.3\% & 96.3\%  \\				
			\cline{2-6}
			& \% False Positive & 1.0\% & 1.0\% & 1.0\% & 1.0\%	\\	
			\hline
		\end{tabular}
		\label{table:cpa_results}
	\end{center}
\end{table}

The major limitation of BNC is its intrinsic function as anomaly detector: indeed, an anomaly is not necessarily malicious. For instance, in a NDN setting, a sudden demand for a video from legitimate users could lead to a temporary increase in traffic, indicating an abnormal activity. To mitigate this problem, \montimage{} considers four possible `states': \textit{normal state}, \textit{IFA attack state}, \textit{CPA attack state}, \textit{number of users increase}. Each state is denoted by different `anomalous' combinations taking into account a total of 18 metrics: a similar solution allows to maintain the FPR to acceptable levels (as shown in Table~\ref{table:cpa_results}). 
We take this opportunity to make a crucial remark for real ML deployments: one may believe that defining more `states' and/or increasing the amount of considered metrics leads to better results. However, according to \montimage{} a similar approach can yield proficient results only in a lab environment because it induces overfitting, and the true \textit{deployment} performance may suffer excessive FPR.

Finally, an intriguing future development of such ML solution involves the consideration of `stateful' analyses that take into account the time-axis (as done, e.g., in~\cite{corsini2021temporal}) and allow to detect even anomalies occurring in the temporal domain. The next case-study by \sgrupo{} will consider a similar application.

\subsection{Combining ML with non-ML methods to protect Industry 4.0 environments}
\label{app:s2grupo}

With the rapid growth of the Industry 4.0 paradigm, industrial environments are even more exposed to Advanced Persistent Threats (APT)~\cite{pierazzi2017online}. Specifically, recent developments of Industrial Control Systems (ICS) represent an attractive target for attackers~\cite{feng2017multi}.
In this case study, we share the experience in the design and operation of \caiac{}\footnote{\url{https://s2grupo.es/en/research-development-innovation/industrial-cybersecurity/caiac.html}}, a non-intrusive device that leverages sequential ML to protect ICS against APT and other cyber-threats.


\paragraph{Scenario and Challenges}
This case study highlights the advantages of ML applications for anomaly detection in time-series data. The intuition is that APT leverage zero-day vulnerabilities, and hence cannot be detected via misuse-based detection approaches---irrespective of being human- or data-driven.
However, point-wise and static anomaly detection approaches are not enough to detect advanced cyberattacks, and the additional perspective provided by the temporal domain allows may facilitate the detection of refined offensive strategies~\cite{pierazzi2017online}. 

In the specific ICS scenario, there are two crucial requirements that must be met by security systems. First, they should operate in a non-intrusive way, avoiding additional overhead and ensuring the regular functionalities of the ICS: this is a tough requirement because ICS include hundreds of devices and while excessive false alarms are annoying, slow reaction times may imply a fallout of the entire ICS. Secondly, they must take into account the complexity and variability of the data in ICS, which is difficult to manage to the intrinsic heterogeneity of ICS. Such requirement cannot be met just with traditional approaches for time-series anomaly detection based on heuristics: to address this problem, \sgrupo{} leverages the capabilities of deep learning.

\paragraph{\sgrupo{} ML-Solution}
The ML solution developed by \sgrupo{}, \caiac{}, is an intriguing example of \textit{ML orchestration} (§\ref{ssec:engineers}): \caiac{} not only leverages the benefits provided by `small' ML models (as done in §\ref{app:montimage}), but also exploits the potential of non-ML methods for time-series analyses. In particular, the idea is to combine deep learning algorithms, epitomized by Long-Short Term Memory (LSTM) neural networks, with statistical approaches for time-series forecasting, such as SARIMA (Seasonal Autoregressive Integrated Moving Average). 
The result is an ensemble of ML and non-ML models, exploiting the benefits of both approaches and overcoming their limitations: statistical models can be more manageable, but when the data has high complexity deep learning is superior.
Such design choice is particularly suited for real ICS deployments due to a threefold advantage with respect to `one-size-fits-all' ML architectures. Specifically:
\begin{itemize}
    \item individual ML models are easier to train because they must deal only with a tiny portion of the data, resulting in better performance and lower false alarms;
    \item it allows combining different algorithms, each addressed to a specific problem and data-type.
    \item it makes the resulting system more `future-proof', because it each ML model can be individually updated, removed, or replaced. 
\end{itemize}
Furthermore, \caiac{} is based on \textit{passive} monitoring in near real-time, hence preventing excessive information overhead while still allowing timely responses.

Let us explain \caiac{} in more detail. The intuition is to analyze the network traffic of the considered ICS from different perspectives, each associated to a specific time-series. Such time-series can differ on the basis of two criteria: the network \textit{metric} (e.g., transmitted packets), and the \textit{granularity} used to aggregate the corresponding metric in time slots of fixed length. 
All such time-series are used to devise multiple ML and non-ML models: the performance of each model can be assessed individually by forwarding its detected anomalies to a higher level correlation layer (similar to~\cite{pierazzi2017online}). The goal of this layer is determining the nature of such anomalies: they can either be legitimate (i.e., a `normal' malfunctioning of a component that must be investigated) or illegitimate (i.e., an attack is taking place). Such procedure allows to identify the most suitable models that will be integrated in \caiac{}, depending on the pros and cons of each model. Indeed, LSTM models may yield a superior performance but require a training phase, whereas statistical models are easier to develop and only require some tuning. Hence, such (non-ML) models are the preferred choice when they exhibit similar performance to LSTM.

\paragraph{Evaluation and Results}
To develop \caiac{}, it is necessary to first assess the characteristics of the specific ICS: indeed, it is not possible to use models trained on different environments (as explained in §\ref{ssec:cots}). Hence, \sgrupo{} monitors and collects the network traffic of the considered ICS, and creates multiple time series, each considering a given metric and granularity.
Some metrics are commonly adopted in NIDS (e.g., transmitted packets or bytes, in-/out-degree~\cite{pierazzi2017online}); others are specific of ICS and require dedicated industrial dissectors that extract the relevant information (e.g., protocol, parameters, command density). Finally, each metric is aggregated in time slots of varying length, from 1 minute to 1 hour.

After this data collection phase, which in the considered setting typically amounts to about 10GB of data per day, \sgrupo{} performs the exploratory analysis focused on determining the most proficient (ML and non-ML) algorithms for studying each time-series. Let us elucidate the differences between two specific applications of SARIMA and LSTM, starting from the non-ML algorithm. 

\begin{figure}[!ht]
\centering
	\includegraphics[width=0.7\textwidth]{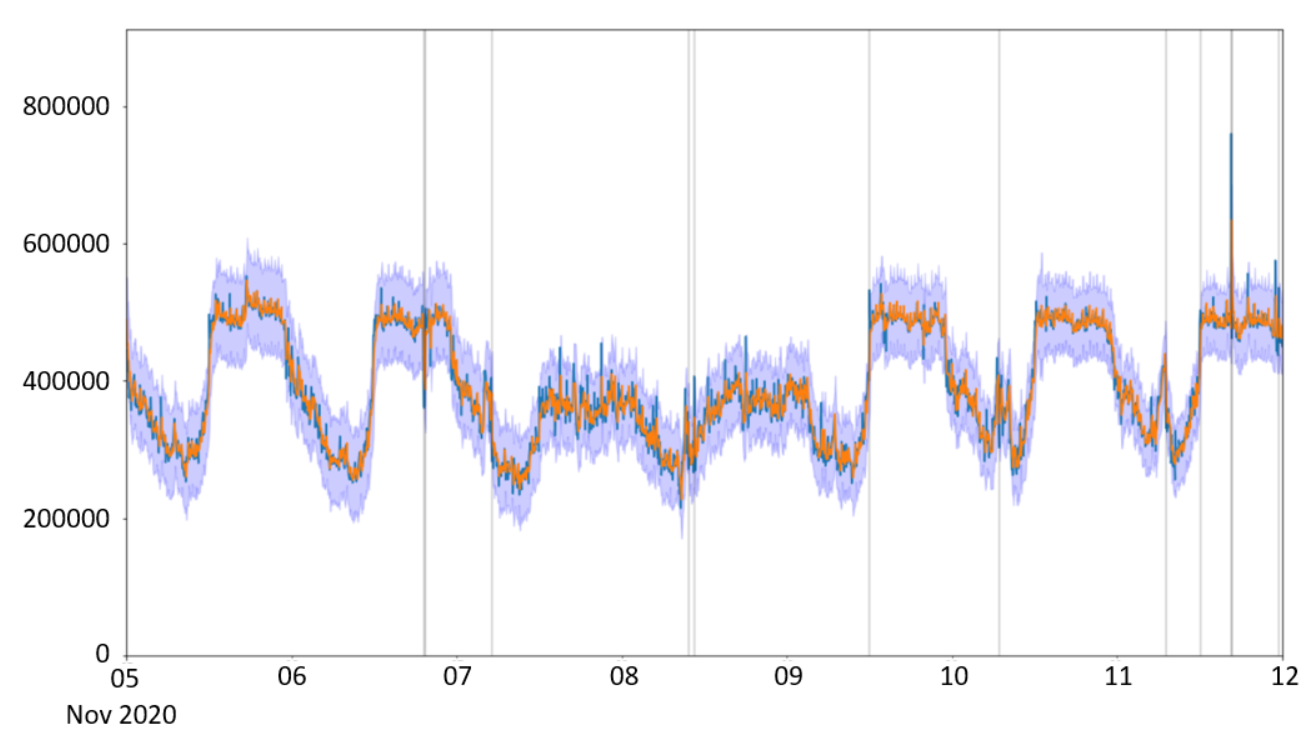}
	\caption{Anomaly detection with (non-ML) SARIMA, using a sliding window of 30 minutes. The time-series represents the transmitted packets (y-axis) within 5 minute slots, over a period of 1 week (x-axis), corresponding to a total of 2K samples. Dark blue correspond to actual values, orange denotes the values predicted with SARIMA, and light blue denotes the confidence interval of SARIMA's predictions. Vertical gray lines correspond to the anomalies detected by SARIMA.}
	\label{fig:sarima}
\end{figure}

Specifically, SARIMA analyzes a time-series by adopting a sliding window approach: all data points within a given time window are considered by SARIMA to predict a `future' value, which is provided alongside a \textit{confidence range}. We provide an example of SARIMA in Fig.~\ref{fig:sarima}, showing the time series of the \textit{transmitted packets} aggregated in time slots of \textit{5 minutes}, over a period of 1 week; the sliding window considered by SARIMA is of 30 minutes. The actual values are reported in dark-blue, whereas the values predicted via SARIMA are shown in orange; the confidence window of each predicted value is shown in light-blue: therefore, actual values that fall outside of such range are treated as anomalous. In particular, vertical gray lines denote the anomalies detected by SARIMA. 

From~\ref{fig:sarima}, we observe that SARIMA accurately detects \textit{stationary} deviations. However, SARIMA can only detect \textit{non-stationary} changes when they happen within its sliding window. Furthermore, non-stationary (but legitimate) changes that occur after a long stationary interval are falsely detected as anomalies by SARIMA. Despite some incorrect predictions, the considered application of SARIMA obtained a performance that was deemed appropriate for the given task, and integrated in \caiac{}.

\begin{figure}[!ht]
\centering
	\includegraphics[width=0.7\textwidth]{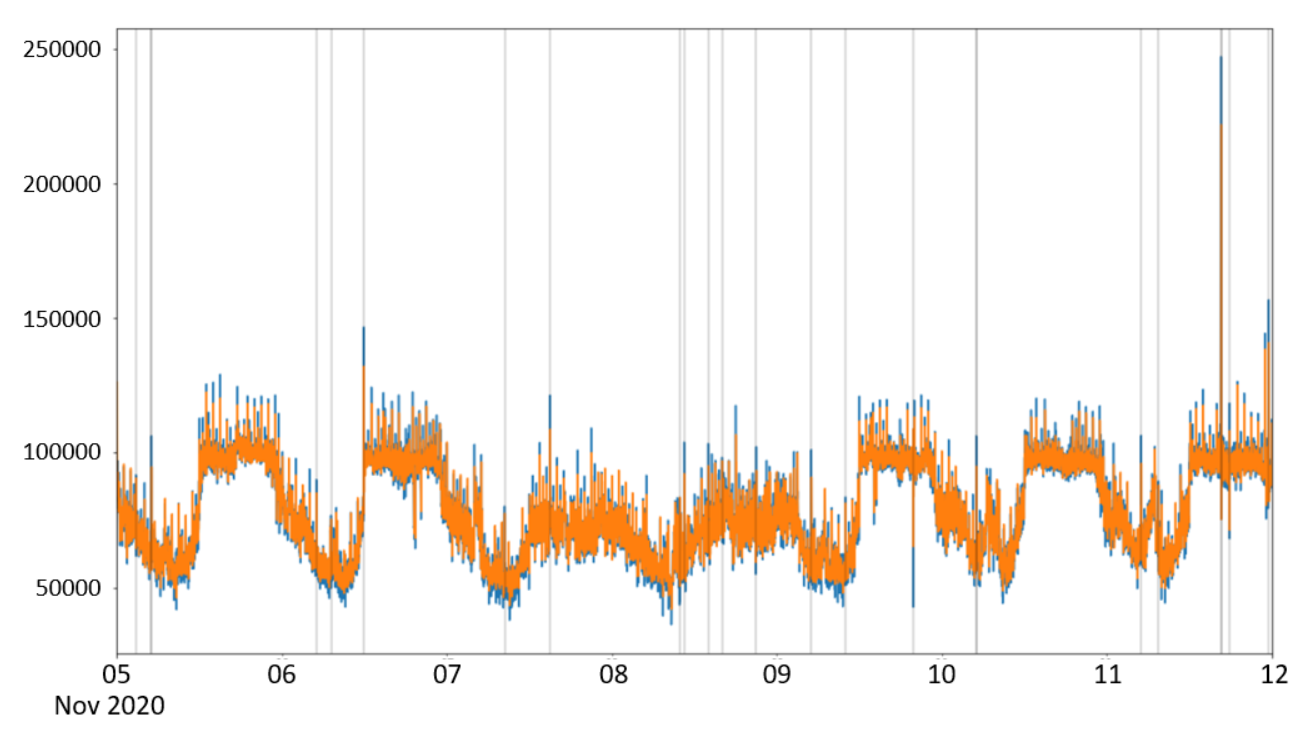}
	\caption{Anomaly detection with a deep LSTM neural network. The time-series represents the transmitted packets (y-axis) within 1 minute slots, over the period of 1 week (x-axis), corresponding to a total of 10K samples. Actual values are shown in blue, and the LSTM predictions are shown in orange. Vertical gray lines denote the anomalies detected by the LSTM.}
	\label{fig:lstm}
\end{figure}

Let us showcase an application of deep learning via LSTM. Since LSTM do not provide a confidence interval for each prediction, \sgrupo{} developed a custom anomaly threshold that takes into account the deviation between predicted and actual values, as well as the degree of accumulation of such deviation in the past history. An example of such LSTM application is given in Fig.~\ref{fig:lstm}, showing the time-series of the transmitted packets (same as Fig.~\ref{fig:sarima}), but with a time slot of 1 minute. The actual values are shown in blue, whereas the LSTM predictions are in orange. Vertical grey lines denote the anomalies detected by the LSTM, i.e., when the actual values falls outside the given anomalous threshold predicted with the LSTM.

From Fig.~\ref{fig:lstm}, we can observe that, by reducing the time slot from 5 to 1 minute, the resulting time-series is less predictable, making statistical methods unfeasible and requiring the advanced capabilities of deep learning. Indeed, the considered LSTM can detect anomalous values without being affected by non-stationary changes---even after long stationary intervals. This example highlights the capabilities of (deep) ML to deal with data with high dimensionality: the LSTM takes into account a long `past' history, allowing to better infer the `normal' behaviour. In contrast, applying SARIMA on the same time-series resulted in very poor results due to the intrinsic variability of the sequence, which forced us to aggregate data in 5 minute time slots.

However, it is important to take into account that the LSTM require a training step, whereas SARIMA only requires some parameter adjustment. In this use-case, the LSTM in Fig.~\ref{fig:lstm} was trained with data collected over \textit{three weeks}. Such characteristic implies that a similar LSTM model requires at least 3 weeks of data collection since no previous network traffic data was available to train the model---alongside the additional computational resources to store such data, and train the LSTM model (which were within acceptable levels). 
Hence, \caiac{} would initially make use of SARIMA, and then replace it after enough data has been collected to develop a more proficient LSTM model.

We can conclude that machine (and deep) learning are powerful instruments for protecting modern ICS, but methods that do not leverage ML are equally important to compensate some of the limitations of ML. As such, future developments should not exclusively focus on ML and overlook the benefits provided by other data-driven methods.

\section{Conclusion}
\label{sec:conclusions}

This paper elucidates the role of Machine Learning (ML) for Cybersecurity by providing a broad and high-level overview of the benefits, problems, and future challenges of ML in this domain. Our paper is oriented at the entire cybersecurity sphere, and to make our contribution understandable by a broad audience we limit technical terms to a minimum. Moreover, we also clarify many misconceptions (summarized in Table~\ref{tab:misconceptions}) that are becoming common due to the increasing abundance of works that link ML with cybersecurity applications.

After introducing the basic concepts of ML, we provide a concise summary of their applications to detect three types of cyber threats: Malware, Phishing, and Network Intrusions. 
Then, we elucidate some additional cybersecurity areas that can leverage the autonomous capabilities of ML, such as raw-data analysis, alert management, cyber risk estimation, and threat intelligence. What follows is a description of the fundamental problems affecting ML within the specific context of operational cybersecurity, which should be known to weigh the pros-and-cons of the still emerging ML solutions. Some of these problems stem from the intrinsic conflicts between the fundamental principles of ML and the cybersecurity domain, and can be addressed only by the joint effort of different worlds: regulatory and authoritative bodies, corporate executives and engineers, as well as the entire scientific community. To this end, we highlight the future challenges of ML in cybersecurity, which we integrate by comprehensive recommendations addressed at each of these separate worlds.
Finally, we present two case studies of successful---and operational---industrial deployments of ML to counter cyber threats.

This paper will hopefully inspire meaningful developments of ML in the cybersecurity domain, laying the foundations for an increased deployment of ML solutions to protect current and future systems.

\begin{table}[!htbp]
    \centering
    \caption{Summary of security ML misconceptions discussed in the paper.}
    \label{tab:misconceptions}
    \resizebox{0.5\columnwidth}{!}{
        \begin{tabular}{c|l|c}
             \toprule
             \textsc{\#} &~~\textsc{Misconception} & \textsc{Ref.}\\
             \toprule
             1 & Deep Learning vs Shallow Learning & §\ref{ssec:glossary} \\
             2 & Machine Learning and Anomaly Detection & §\ref{sec:detection} \\
             3 & Legitimacy of Adversarial Samples & §\ref{ssec:general} \\
             4 & Minimal Adversarial Perturbations & §\ref{ssec:general} \\
             5 & Size of training data & §\ref{ssec:inhouse} \\
             6 & Updating ML models with new data & §\ref{ssec:inhouse} \\
             
             \bottomrule

        \end{tabular}
    }
\end{table}



\end{document}